\documentclass[aps,10pt,amsmath,amssymb,titlepage,prb,twocolumn,superscriptaddress,floatfix]{revtex4-2} %added floatfix for revtex
\usepackage{import}
\usepackage{amsfonts}
\usepackage{braket}
\usepackage{times}
\usepackage{graphicx}
\usepackage{color}
\usepackage{fontawesome5}
\usepackage{figchild}
\usepackage[normalem]{ulem}
\graphicspath{{Figures/}}
\usepackage{siunitx}
\usepackage[caption=false]{subfig}
\usepackage[section]{placeins}
\usepackage{xspace} 
\usepackage{url} 

\usepackage{breakurl}
\usepackage[colorlinks=true, allcolors=blue,breaklinks]{hyperref}
\usepackage{capt-of}
\usepackage[switch]{lineno}  
\usepackage{mathtools}
\usepackage{mathrsfs}
\usepackage{textcomp}%
\usepackage{booktabs}%
\usepackage{algorithm}%
\usepackage{algorithmicx}%
\usepackage{algpseudocode}%
\usepackage{listings}%
\usepackage{bm}
\usepackage{commath}
\makeatletter
\def\maketitle{
\@author@finish
\title@column\titleblock@produce
\suppressfloats[t]}
\makeatother

\usepackage{xpatch}

\makeatletter
\patchcmd{\@ssect@ltx}
    {\addcontentsline{toc}{#1}{\protect\numberline{}#8}}
    {}
    {}
    {}
\makeatother

\RequirePackage[T1]{fontenc}
\RequirePackage[utf8]{inputenc}
\RequirePackage{lmodern}

\usepackage[nomain, acronym]{glossaries}
\DeclareCaptionLabelSeparator{bar}{ | }
\newcommand{\Sb}{$^{123}$Sb}

\renewcommand{\Vec}[1]{\boldsymbol{#1}}          % physicist vector, greek
\newcommand{\ketbra}[2]{\ket{#1}\!\!\bra{#2}}

\newcommand{\proj}[1]{\ketbra{#1}{#1}}
\newcommand{\Tr}{\mathrm{Tr}}
\newcommand{\beginsupplement}{%
        \setcounter{table}{0}
        \renewcommand{\thetable}{S\arabic{table}}%
        \setcounter{figure}{0}
        \renewcommand{\thefigure}{S\arabic{figure}}%
        \setcounter{section}{0}
        \renewcommand*{\thesection}{SI:~\arabic{section}}
     }

\begin{document} % moved here for revtex 

\title{Schr\"{o}dinger cat states of a nuclear spin qudit in silicon}

\author{Xi Yu}
\affiliation{School of Electrical Engineering and Telecommunications, UNSW Sydney, Sydney, NSW 2052, Australia}
\affiliation{Centre for Quantum Computation and Communication Technology}
\author{Benjamin Wilhelm}
\affiliation{School of Electrical Engineering and Telecommunications, UNSW Sydney, Sydney, NSW 2052, Australia}
\affiliation{Centre for Quantum Computation and Communication Technology}
\author{Danielle Holmes}
\affiliation{School of Electrical Engineering and Telecommunications, UNSW Sydney, Sydney, NSW 2052, Australia}
\affiliation{Centre for Quantum Computation and Communication Technology}
\author{Arjen Vaartjes}
\affiliation{School of Electrical Engineering and Telecommunications, UNSW Sydney, Sydney, NSW 2052, Australia}
\affiliation{Centre for Quantum Computation and Communication Technology}
\author{Daniel Schwienbacher}
\affiliation{School of Electrical Engineering and Telecommunications, UNSW Sydney, Sydney, NSW 2052, Australia}
\affiliation{Centre for Quantum Computation and Communication Technology}
\author{Martin Nurizzo}
\affiliation{School of Electrical Engineering and Telecommunications, UNSW Sydney, Sydney, NSW 2052, Australia}
\affiliation{Centre for Quantum Computation and Communication Technology}
\author{Anders Kringh\o{}j}
\affiliation{School of Electrical Engineering and Telecommunications, UNSW Sydney, Sydney, NSW 2052, Australia}
\affiliation{Centre for Quantum Computation and Communication Technology}
\author{Mark R. van Blankenstein}
\affiliation{School of Electrical Engineering and Telecommunications, UNSW Sydney, Sydney, NSW 2052, Australia}
\affiliation{Centre for Quantum Computation and Communication Technology}
\author{Alexander M. Jakob}
\affiliation{School of Physics, University of Melbourne, Melbourne, VIC 3010, Australia}
\affiliation{Centre for Quantum Computation and Communication Technology}
\author{Pragati Gupta}
\affiliation{Institute for Quantum Science and Technology, University of Calgary, Alberta T3A~0E1, Canada}
\author{Fay E. Hudson}
\affiliation{School of Electrical Engineering and Telecommunications, UNSW Sydney, Sydney, NSW 2052, Australia}
\affiliation{Diraq Pty. Ltd., Sydney, NSW, Australia}
\author{Kohei M. Itoh}
\affiliation{School of Fundamental Science and Technology, Keio University, Kohoku-ku, Yokohama, Japan}
\author{Riley J. Murray}
\affiliation{Quantum Performance Laboratory, Sandia National Laboratories, Albuquerque, NM 87185, USA}
\author{Robin Blume-Kohout}
\affiliation{Quantum Performance Laboratory, Sandia National Laboratories, Albuquerque, NM 87185, USA}
\author{Thaddeus D. Ladd}
\affiliation{HRL Laboratories, LLC, Malibu, CA 90265-4797 USA}
\author{Namit Anand}
\affiliation{Quantum Artificial Intelligence Laboratory (QuAIL), NASA Ames Research Center, Moffett Field, CA, 94035, USA}
\affiliation{KBR, Inc., 601 Jefferson St., Houston, TX 77002, USA}
\author{Andrew S. Dzurak}
\affiliation{School of Electrical Engineering and Telecommunications, UNSW Sydney, Sydney, NSW 2052, Australia}
\affiliation{Diraq Pty. Ltd., Sydney, NSW, Australia}
\author{Barry C. Sanders}
\affiliation{Institute for Quantum Science and Technology, University of Calgary, Alberta T3A~0E1, Canada}
\author{David N. Jamieson}
\affiliation{School of Physics, University of Melbourne, Melbourne, VIC 3010, Australia}
\affiliation{Centre for Quantum Computation and Communication Technology}
\author{Andrea Morello}
\thanks{To whom correspondence should be addressed; E-mail: a.morello@unsw.edu.au}
\affiliation{School of Electrical Engineering and Telecommunications, UNSW Sydney, Sydney, NSW 2052, Australia}
\affiliation{Centre for Quantum Computation and Communication Technology}

\begin{abstract}
    High-dimensional quantum systems are a valuable resource for quantum information processing. They can be used to encode error-correctable logical qubits, which has been demonstrated using continuous-variable states in microwave cavities or the motional modes of trapped ions. For example, high-dimensional systems can be used to realise `Schr\"{o}dinger cat' states, superpositions of widely displaced coherent states that can also be used to illustrate quantum effects at large scales. Recent proposals have suggested encoding qubits in high-spin atomic nuclei, finite-dimensional systems that can host hardware-efficient versions of continuous-variable codes. Here we demonstrate the creation and manipulation of Schrodinger cat states using the spin-7/2 nucleus of an antimony atom embedded in a silicon nanoelectronic device. We use a multi-frequency control scheme to produce spin rotations that preserve the symmetry of the qudit, and constitute logical Pauli operations for qubits encoded in the Schrodinger cat states. Our work demonstrates the ability to prepare and control nonclassical resource states, a prerequisite for applications in quantum information processing and quantum error correction using our scalable, manufacturable semiconductor platform. 
    
\end{abstract}

% \date{}% removed for revtex 

\maketitle

\ \\
A spin-1/2 particle is the textbook example of physical object in which to encode one qubit (Hilbert space dimension $d=2$) in a discrete-variable paradigm. The `quantumness' of a spin was recognized since the Stern-Gerlach experiment, but is otherwise surprisingly elusive. The dynamics of a spin-1/2 maps directly to the precession of a classical gyroscope \cite{slichter2013principles}, and its statistics can be cast within local hidden-variable models \cite{kochen1967problem}. In the language of phase-space representations, the quasi-classicality of spin-1/2 particles is captured by the fact that they can only support minimum-uncertainty, Gaussian-like spin coherent states \cite{kitagawa1993squeezed}. Entangling multiple qubits to obtain a $d\gg 2$-dimensional Hilbert space is therefore essential to capture the true power of quantum information.

At the other extreme, continuous-variable quantum computing encodes information in the intrinsically $d~\rightarrow~\infty$ dimensional Hilbert space of a quantum harmonic oscillator \cite{gottesman2001encoding,braunstein2005quantum,fluhmann2019encoding,campagne2020quantum}. The state of the system is described by the complex quadratures of a bosonic field, and the `quantumness' of the encoded states becomes rather transparent. Classical fields display a Gaussian-shaped distribution in the complex plane, whereas quantum resource states have non-Gaussian statistics \cite{walschaers2021non}. A famous example is the Schr\"{o}dinger cat state, a superposition of two coherent states far displaced from each other \cite{ourjoumtsev2007generation,grimm2020stabilization,lescanne2020exponential}. Its quantumness is captured by the negativity of the Wigner function \cite{ferrie2011quasi}, which also implies contextuality and, for multi-particle systems with spatial extent, non-locality. 
The key to the creation of nonclassical states is always a nonlinearity, without which it would only be possible to displace trivial Gaussian states.

The intermediate regime, comprising qu\textit{d}its with $3 \leq d < \infty$, is very rich \cite{wang2020qudits,anderson2015accurate,blok2021quantum,chalopin2018quantum,hrmo2023native}, but relatively unexplored. Once $d\geq 3$, it becomes possible to create states that violate local hidden-variable theories \cite{kochen1967problem}, and to encode error-correctable logical qubits within a single quantum object \cite{gross2021designing,gross2021hardware,omanakuttan2024fault}, without resorting to entangling multiple physical qubits. 

In this work we experimentally demonstrate the creation and manipulation of Schr\"{o}dinger cat states of a single antimony-123 (\Sb) nuclear spin qudit in silicon \cite{asaad2020coherent,fernandez2024navigating}. The $I=7/2$ nuclear spin of \Sb~results in a $d=2I+1=8$ dimensional Hilbert space. Although 8 is not infinity, it is large enough to accommodate nontrivial quantum states with properties remarkably similar to bosonic states in continuous-variable systems (Supplementary Information, Sections \ref{sec:cat_code} and \ref{representations}). Crucially, it is enough to support a logical qubit encoding capable of correcting all first-order rotation errors \cite{gross2021designing}.

\begin{figure}[ht!]

\includegraphics[width=\columnwidth]{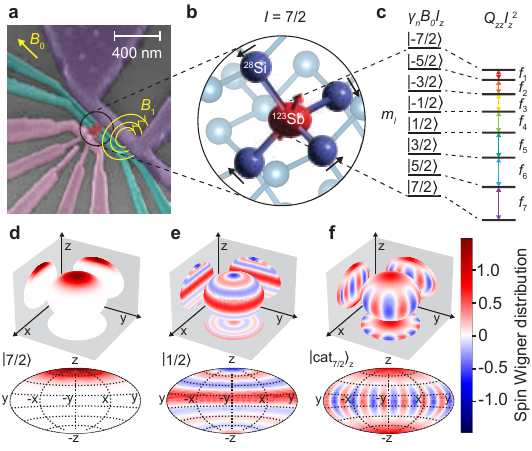}
    \caption{
    \textbf{The 8-dimensional $^{123}$Sb nuclear spin qudit in silicon.}
    \textbf{a}, False-colour scanning electron micrograph of a characteristic silicon device showing the single electron transistor (pink), gates for tuning the electric potential of the donor (green), microwave antenna (purple) and donor implant window (red). The donor is subject to an oscillating magnetic field $B_1$ from the microwave antenna which is perpendicular to an in-plane external field $B_0=1.384$~T. \textbf{b}, The $^{123}$Sb donor in the $^{28}$Si crystal lattice. Black arrows illustrate the shear strain on Si atoms bonded to $^{123}$Sb which creates an electric field gradient, resulting in a nuclear quadrupole shift. \textbf{c}, Energy levels of the ionised $^{123}$Sb donor. The Zeeman energy, $\gamma_nB_0\hat{I}_z$, produces equispaced nuclear energy levels, while the quadrupole coupling, here in the simplified form $Q_{zz}\hat{I}_z^2$, shifts the energy levels depending on spin projection $m_I$, resulting in 7 individually addressable NMR transitions, labelled by the coloured arrows. \textbf{d,e,f}, Theoretical spin Wigner function on a sphere (top) \cite{millen2023phasespace} and corresponding Hammer projections (bottom) for the example states: \textbf{d} $\ket{7/2}$ and \textbf{e} $\ket{1/2}$ spin projection eigenstates, and \textbf{f} $\ket{\text{cat}_{7/2}}_z$ Schr\"{o}dinger cat state. The grey planes in the Wigner plots show the mirrored reflections of the sphere. The colour bar for Wigner function shown here is used without rescaling throughout the rest of this work.}
    \label{fig:figure_1}
\end{figure}

\subsection*{SU(8) and SU(2) operations in a generalised rotating frame}
The device structure follows that of early experiments on single donors in silicon (Fig.~\ref{fig:figure_1}a). During the quantum operations, the \Sb~donor is kept in the ionised, charge-positive $D^+$ state. In the absence of a hyperfine-coupled electron, the nuclear spin Hamiltonian (in frequency units) takes the form:
\begin{equation}
\label{eq:H_ion}
\hat{\mathcal{H}}_{D^{+}} = -\gamma_{\rm n}B_{0}\hat{I}_z+\sum_{\alpha,\beta\in\{x,y,z\}} Q_\mathrm{\alpha\beta} \hat{I}_\mathrm{\alpha}\hat{I}_\mathrm{\beta},
\end{equation}
where $B_0 =1.384$~T is a static magnetic field, $\gamma_{\rm n}=5.55$~MHz/T is the nuclear gyromagnetic ratio, $\alpha,\beta=\{x,y,z\}$ are Cartesian axes, $\hat{I}_{\alpha}$ are the 8-dimensional nuclear spin  operators,
and $Q_\mathrm{\alpha\beta}$ is the interaction energy between the electric quadrupole moment of the nucleus and an electric field gradient, which arises mostly from local strain breaking the cubic symmetry of the silicon lattice \cite{asaad2020coherent} (Fig.~\ref{fig:figure_1}b). 
 
The electric quadrupole interaction is the nonlinear term that enables universal control of the antimony qudit. While the Zeeman interaction, a linear function of the $\hat{I}_z$ operator, splits the energies of the nuclear spin eigenstates $\ket{I,m_{I}}, m_I = -I, -I+1, \ldots I$ uniformly by $f_0^+ = \gamma_{\rm n}B_0 \approx 7.7$~MHz (the superscript $^+$ labels quantities pertaining the ionised $D^+$ donor state), the quadrupole term introduces a non-uniform spacing, so that the nuclear magnetic resonance (NMR) frequencies between pairs of eigenstates differ by $f_{\rm q}^+ \approx 28$~kHz (Fig.~\ref{fig:figure_1}c). Since $f_0^+ \gg f_{\rm q}^+$, the eigenstates of $\hat{\mathcal{H}}_{D^{+}}$ are simply the $\ket{m_I}$ eigenstates of $\hat{I_z}$ (in the following we drop the spin quantum number $I$ from the notation, since it is always $7/2$). Notably, only the $\ket{\pm 7/2}$ states (Fig.~\ref{fig:figure_1}d) are spin coherent states, whereas all other eigenstates of $\hat{I}_z$ have non-Gaussian Wigner functions which display strong negativity (Fig.~\ref{fig:figure_1}e).
 
This energy level structure allows for universal control of the spin qudit \cite{fernandez2024navigating}, by applying a nuclear magnetic resonance (NMR) control Hamiltonian of the form:
\begin{equation}
  \hat{\mathcal{H}}_\mathrm{1}(t) = -\gamma_\mathrm{n} \hat{I}_x \sum_{k=1}^{2I}\cos(2\pi f_k t + \phi_k)B_{1,k}(t),
  \label{eq:H_1}
\end{equation}
where $f_k$ are the NMR frequencies ($f_1 = \langle -7/2|\hat{\mathcal{H}}_{D^{+}}| -7/2\rangle - \langle -5/2|\hat{\mathcal{H}}_{D^{+}}| -5/2\rangle$, etc.), $B_{1,k}(t)$ the oscillating magnetic field amplitudes, and $\phi_k$ the phases. The 7 amplitudes $B_{1,k}(t)$ and 7 phases $\phi_k$ provide the $(d^2 -1)-(d-1)^2=14$ independent parameters required for arbitrary SU(8) state generation (recognising a U(7) equivalence of states). The time-dependence of the pulses allows universal control for generating any SU($d$) unitary in $\mathcal O(d^2)$ steps~\cite{reck1994experimental,bullock2005}. 
We produce $\hat{\mathcal{H}}_\mathrm{1}(t)$ by direct digital synthesis with an FPGA waveform generator. Crucially, the generator's software creates 7 `virtual clocks' that allow us to define a generalised rotating frame \cite{leuenberger2003grover} (GRF - see Supplementary Information, Section~\ref{sec:GlobalRotationTheory}) wherein the spin state appears static in the absence of drives. In other words, the GRF cancels all terms of the static Hamiltonian, whereas a single-frequency frame would only cancel terms $\propto \hat{I}_z$. The phases $\phi_k$ of the virtual clocks can be shifted by software instructions alone, allowing the establishment of arbitrary relative phases between the 8 levels, corresponding to $\hat{I}_z$ rotations, $\hat{I}_z^2$ rotations, and in fact any diagonal unitary,  whereas a single-frequency rotating frame and frame shifting would only enact $\hat{I}_z$ control.

\begin{figure}[ht!]
    \centering
\includegraphics[width=\columnwidth]{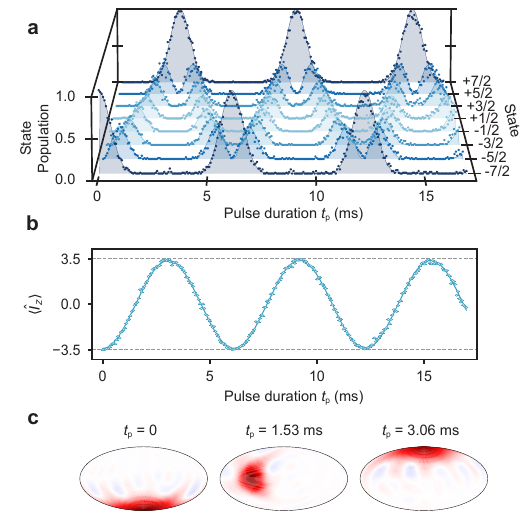}
	 \caption{\textbf{Covariant SU(2) rotations.} \textbf{a}, Population of the $\ket{m_I}$ states as a function of duration $t_{\rm p}$ of an NMR pulse of 7 tones with equal amplitude, driving a covariant SU(2) rotation of an initial $\ket{-7/2}$ spin coherent state. Solid lines are calculations (not fits) assuming perfect initialisation and control. Here and elsewhere, data points are raw values, with no post-selection or state preparation and measurement (SPAM) error extraction. \textbf{b}, Expectation value of $\hat{I}_z$ as a function of pulse duration, calculated from the data in \textbf{a}. \textbf{c}, Wigner function of reconstructed states with rotation angle $\Theta = 0, \pi/2, \pi$ at the times indicated by $t_{\rm p}$, showing that the Wigner distribution is rotated rigidly around the $-y$ axis.
    }
    \label{fig:figure_2}
\end{figure}
 
Within the GRF, we demonstrate a covariant  rotation (CR) \cite{faist2020continuous} of the large spin, i.e.~an SU(2) rotation where the spin state is rigidly rotated around an axis of the Bloch sphere, preserving the shape of the Wigner function. We first prepare the spin in the $\ket{-7/2}$ eigenstate (see Supplementary Information, Section~\ref{sec:nuclear_readout}, for spin initialisation), then simultaneously apply 7 tones at $\{f_k\}$, all with equal amplitudes (22.86~mV for each tone), resulting in equal magnetic field amplitudes $B_{1,k}(t) = B_1(t)$ for all $k$. Since the tones are applied exactly on-resonance with each $f_k$, the rotations are always around an axis placed on the equator of the GRF, denoted by a single longitude $\varphi$. Calling $\Theta$ the rotation angle, we will denote such operations with $R_{\Theta}(\varphi)$. Monitoring the populations of the $\ket{m_I}$ eigenstates as a function of pulse duration, i.e. $\Theta$ (Fig.~\ref{fig:figure_2}a), we find the predicted behaviour for a spin undergoing a rotation around the $-y$-axis of the $I=7/2$ Bloch sphere. Calculating the $\langle \hat{I}_z \rangle$ expectation value reveals the expected behaviour of a covariant Rabi oscillation of the large spin (Fig.~\ref{fig:figure_2}b), with Rabi frequency $f_{\rm Rabi}^{\rm CR} = 163.4(1)$~Hz. Here and elsewhere, error bars represent $1\sigma$ confidence intervals. Snapshots of the spin Wigner function (see Supplementary Information, Section~\ref{representations}, for spin Wigner function, and Section~\ref{Density matrix reconstruction}, for density matrix reconstruction) at different times confirm the picture of a smoothly evolving spin coherent state, preserving its shape as it rotates (Fig.~\ref{fig:figure_2}c,d,e). The experiment was conducted in the $\gamma_\mathrm{n} B_1 \ll f_{\rm q}^+$ regime (see Supplementary Information, Section~\ref{sec:GlobalRotationTheory}, for a discussion of the general case), where the power broadening is negligible compared to the spacing between the NMR resonances, making it sufficient to adopt simple rectangular $B_{1}(t)$ pulse envelopes.

\subsection*{Generation and manipulation of Schr\"{o}dinger cat states}

\begin{figure*}[ht!]
    \centering
\includegraphics[width=\linewidth]{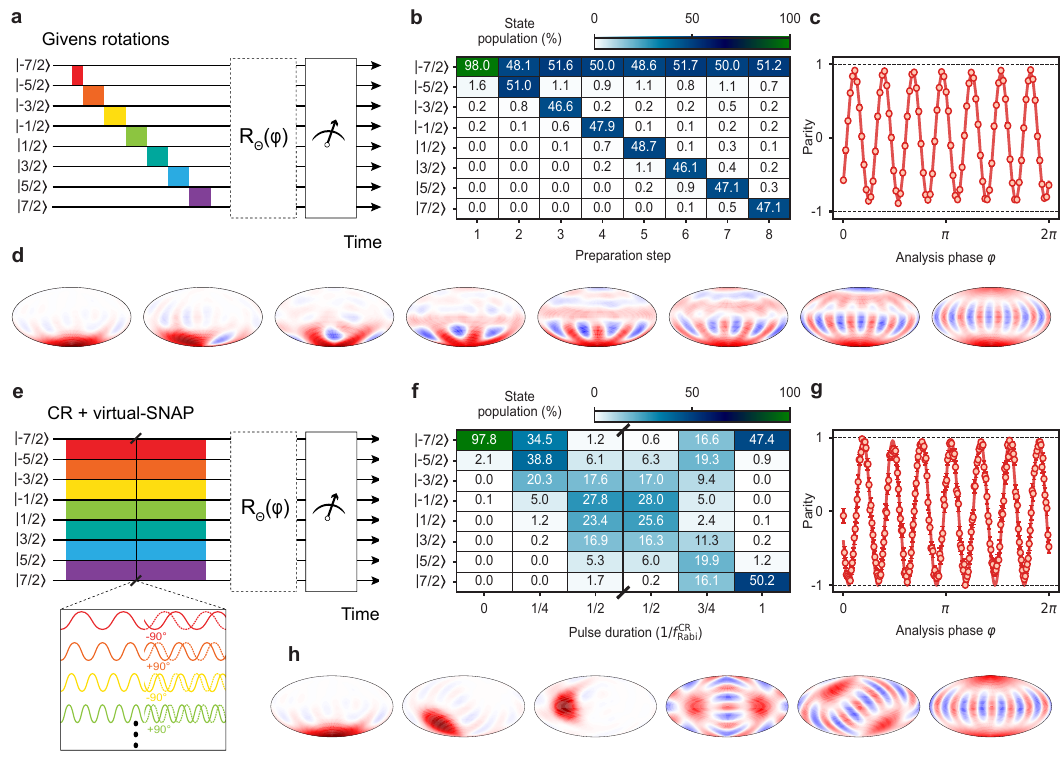}
    \caption{\textbf{Creation of Schr\"{o}dinger cat states}. \textbf{a}, Pulse sequence for cat state creation via Givens rotations, followed by a covariant SU(2) rotation $R_{\Theta}(\varphi)$ for tomography, and measurement of the nuclear state populations. \textbf{b}, State populations at each step. \textbf{c}, Parity of the state after the tomography rotation pulse $R_{\pi/2}(\varphi)$, displaying the expected 7 periods of oscillation, and a contrast $C_{\rm p} = 0.878(3)$ of the cat state $\ket{\rm cat_{7/2}, \xi_7=\pi}_z = \left( \ket{7/2}-\ket{-7/2}\right) / \sqrt{2}$. \textbf{d}, Reconstructed Wigner functions at each step. Note the increasing number of interference fringes. \textbf{e}, Pulse sequence for cat state creation via covariant SU(2) rotations (CR) and virtual-SNAP, implemented by shifting the phases of the generalised rotating frame as per the diagram below. \textbf{f}, State populations at the indicated steps along the protocol. \textbf{g}, Parity of the state after the tomography rotation pulse $R_{\pi/2}(\varphi)$, displaying 7 periods of oscillations and a contrast $C_{\rm p} = 0.982(5)$ of the cat state $\ket{\rm cat_{7/2}, \xi_7=\pi/2}_z = \left( \ket{7/2}+i\ket{-7/2}\right) / \sqrt{2}$. \textbf{h}, Reconstructed Wigner functions at the indicated steps. Note the instantaneous transition from a $-x$-oriented coherent state to a cat state, caused by the virtual-SNAP.}
    \label{fig:figure_3}
\end{figure*}

With SU(8) and SU(2) rotations available, we proceed to generate the $z$-oriented Schr\"{o}dinger cat state of the high-spin nucleus, $\ket{\rm cat_{7/2}}_z = \left( \ket{7/2}+e^{i\xi_7}\ket{-7/2}\right) / \sqrt{2}$ (Fig.~\ref{fig:figure_1}f), using two different methods. The first, based on Givens rotations \cite{cybenko2001reducing} involves simply preparing the $\ket{-7/2}$ state, applying a $\pi / 2$ pulse at $f_1$ to produce $\left( \ket{-7/2}-\ket{-5/2}\right) / \sqrt{2}$, and then a sequence of $\pi$ pulses between ascending pairs of states (Fig.~\ref{fig:figure_3}a,b). The quality of the resulting $\ket{\rm cat_{7/2}}_z$ state can be assessed by measuring the contrast $C_{\rm p}$ of the parity oscillations around the equator of the Bloch sphere (Fig.~\ref{fig:figure_3}c). For this, we apply CRs $R_{\pi/2}(\varphi)$ around different axes in the GRF, indexed by the longitude $\varphi$, then measure the expectation value of the parity operator $\hat{\Pi} =  \sum_{m_I} (-1)^{I+m_I}\ket{m_I}\bra{m_I}$ of the resulting states. The parity oscillations display a contrast $C_{\rm p} = 0.878(3)$ that we extracted via a sinusoidal fit. We also plot the Wigner function (Fig.~\ref{fig:figure_3}d) from the reconstructed density matrix $\rho_{\rm MLE}$, obtained from maximum likelihood estimation quantum state tomography (see Supplementary Information, Section~\ref{Density matrix reconstruction}). The Wigner function reveals the increasing number of interference fringes at each step of the process, from 0 in the $\ket{-7/2}$ state, to 7 in the $\ket{\rm cat_{7/2}}_z$ state. From the reconstructed density matrix $\rho_{\rm MLE}$ we calculate the state fidelity $\mathcal{F} = \braket{\psi|\rho_{\rm MLE}|\psi} = 0.794(2)$, where $\ket{\psi}=\ket{\rm cat_{7/2}}_z = \left( \ket{7/2}+e^{i\xi_7}\ket{-7/2}\right) / \sqrt{2}$ is the target state, with $\xi_7=\pi$. Unlike the contrast of the parity oscillation, the state fidelity depends sensitively on the phase of the state relative to the target. The state reconstruction reveals a systematic phase shift $\Delta{\xi_{7}} = +39.3^{\circ}$, which could in principle be cancelled by a redefinition of the GRF (see below). Removing the phase shift results in $\mathcal{F} = 0.884(2)$ for this cat state.

An alternative method to generate the $\ket{\rm cat_{7/2}}_z$ state \cite{gupta2024robust} makes explicit use of the software-defined GRF. Starting from $\ket{-7/2}$, we apply a $\pi/2$ covariant SU(2) rotation to prepare a coherent state on the equator of the Bloch sphere. We then send an instruction to the FPGA to redefine the phases $\phi_k$ of the GRF, shifting the odd clocks ($f_{1,3,5,7}$) by $-90^{\circ}$ and the even ones ($f_{2,4,6}$) by $+90^{\circ}$. This embodies one-axis twisting dynamics \cite{gupta2024robust}, instantly and error-free (to within the time resolution of the FPGA, 4~ns). In the redefined GRF, the state of the spin is now the $x$-oriented cat state $\ket{\rm cat_{7/2}}_x$ (see Supplementary Information, Section~\ref{SNAP}). We call this operation `virtual-SNAP', since it is mathematically analogous to the Selective Number-dependent Arbitrary Phase (SNAP) operation, first introduced in microwave cavities coupled to superconducting qubits \cite{heeres2015cavity}. In contrast to the linear cavity system, our system does not require an ancilla qubit to achieve state-selective phase shifts, because it possesses an intrinsic non-linearity, in the form of a quadrupole splitting, that permits state-selective operations. The operation is virtual because no physical action is applied to the system; it can be viewed as the multi-level extension of the virtual-$Z$ gate \cite{mckay2017efficient}.

From the $\ket{\rm cat_{7/2}}_x$, a further SU(2) $\pi/2$ rotation creates the cat state $\ket{\rm cat_{7/2}, \xi_7=\pi/2}_z = \left( \ket{7/2}+i\ket{-7/2}\right) / \sqrt{2}$. The cat state produced by virtual-SNAP achieves a contrast of the parity oscillations $C_{\rm p}=0.982(5)$ and a state fidelity $\mathcal{F}=0.874(2)$ from the reconstructed density matrix. Extracting the $\Delta\xi_7 = +24.1^{\circ}$ phase shift between the created cat and the target yields $\mathcal{F}=0.913(2)$. The method of producing cat states by virtual-SNAP plus covariant SU(2) rotations is a key novelty of our work, and appears to yield superior fidelity compared to the Givens rotation method. 

Cat states can be defined in subspaces other than $m_I = \pm 7/2$. For this purpose, we initialise the spin in the $\ket{-5/2}, \ket{-3/2}$, or $\ket{-1/2}$ states, and then use the Givens rotations method to prepare $\ket{\rm cat_{5/2}}_z = \left( \ket{5/2}+e^{i\xi_5}\ket{-5/2}\right) / \sqrt{2}$, $\ket{\rm cat_{3/2}}_z = \left( \ket{3/2}+e^{i\xi_3}\ket{-3/2}\right) / \sqrt{2}$, and the trivial (i.e. non-cat) state $\left( \ket{1/2}+e^{i\xi_1}\ket{-1/2}\right) / \sqrt{2}$. However, in these subspaces it is no longer true that a multi-frequency drive with all equal amplitudes results in a covariant SU(2) rotation, so the values of $B_{1,k}$ must be individually calibrated (see Supplementary Information, Section~\ref{SI_sec:subRot}, for details). 

\subsection*{Coherence times}

\begin{figure}[ht!]
    \centering
\includegraphics[width=\columnwidth]{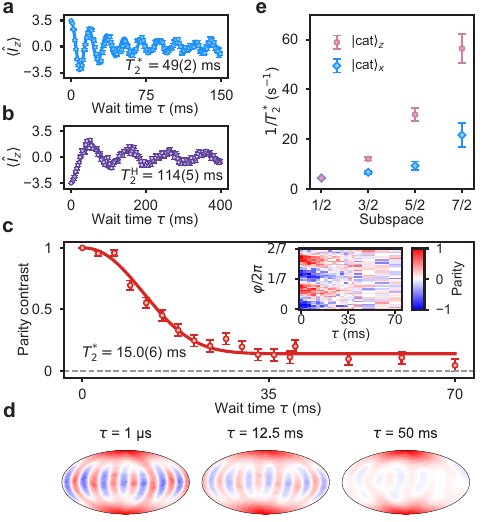}
    \caption{\textbf{Dephasing of spin coherent states and Schr\"{o}dinger cat states.}
    \textbf{a}, Decay of Ramsey fringes for a spin coherent state prepared with a SU(2) $\pi/2$ pulse starting from $\ket{-7/2}$, left to freely evolve for time $\tau$, then rotated with a second SU(2) $\pi/2$ pulse around an angle $\varphi$ increasing with $\tau$. The dephasing time is $T_2^{\ast}=49(2)$~ms. \textbf{b}, Hahn echo experiment, with additional SU(2) $\pi$ pulse at the halfway point, yielding coherence time $T_2^{\rm H}=114(5)$~ms. \textbf{c}, Decay of the parity oscillations of a $\ket{\rm cat_{7/2}}_z$ state with wait time, extracted from the data in the inset. The cat state is prepared by the virtual-SNAP method and has coherence time $T_2^{\ast}=15.0(6)$~ms. \textbf{d}, Reconstructed Wigner functions of an initial $\ket{\rm cat_{7/2}}_z$ after the indicated wait times. Note the disappearance of interference fringes around the equator at long times. \textbf{e}, dephasing rates for $z$-cats ($\circ$) and $x$-cats ($\Diamond$) as a function of subspace size. Here, the subspace cats $\ket{{\rm cat}_{I}}_x$ are prepared by the virtual-SNAP method, and $\ket{{\rm cat}_{I}}_z$ are prepared by Givens rotations (see Methods). Note that the $x$ and $z$ states in the $m_I=1/2$ subspace coincide.}
    \label{fig:figure_4}
\end{figure}

The multi-level control methods shown above represent a promising new direction in quantum information processing. Furthermore, applying them to nuclear spins in isotopically-enriched $^{28}$Si gives access to a platform with exceptionally long coherence times \cite{saeedi2013room,muhonen2014storing}. For single \Sb~nuclei, the only data existing to date quantified the coherence of superpositions of two spin projection states with $\Delta m_I = 1$ or $2$ \cite{asaad2020coherent,fernandez2024navigating}. Here we report the coherence times of spin coherent and Schr\"{o}dinger cat states. Fig.~\ref{fig:figure_4}a reports a `covariant Ramsey' experiment, where we apply a covariant SU(2) $\pi/2$ rotation around $-y$ in the GRF, wait a variable time $\tau$, and then apply a second covariant SU(2) $\pi/2$ rotation around an equatorial axis, shifted from $-y$ by a phase that increases with $\tau$ in order to observe fringes in $\langle \hat{I}_z \rangle(\tau)$ (the pulses are applied exactly on-resonance). We fit the observed fringes with a decaying sinusoid of the form $\langle \hat{I}_z \rangle(\tau) \propto 3.5 \exp[-(\tau/T_2^{\ast})^{\alpha_{\rm R}}]$, where  $T_2^{\ast} = 49(2)$~ms is the dephasing time and $\alpha_{\rm R} = 0.84(4)$ is the exponent of the Ramsey decay. A covariant Hahn echo is obtained by adding a covariant SU(2) $\pi$-pulse in the middle of the sequence (Fig~\ref{fig:figure_4}b). The extracted Hahn echo coherence time (fitting a decay with same the form as the Ramsey) is $T_2^{\rm H} = 114(5)$~ms, with decay exponent $\alpha_{\rm H} = 0.61(3)$. The values of $\alpha_{\rm R,H} < 1$ indicate stretched exponential decays, consistent with the fact that the spin coherent state prepared by the first covariant SU(2) $\pi/2$ pulse is a Gaussian-like superposition of all $\ket{m_I}$ eigenstates, and the coherence between pairs of such states varies substantially depending on $m_I$ (see Supplementary Information, Section~\ref{SI_sec:T2Star}). 
 
For the Schr\"{o}dinger cat states, we measure the contrast $C_{\rm p}$ of the parity oscillations while adding a wait time $\tau$ between preparation and measurement (Fig.~\ref{fig:figure_4}c). For the $\ket{\rm cat_{7/2}}_z$ state we find a dephasing time $T_2^{\ast}=15.0(6)$~ms.
Smaller cat states ($\ket{\rm cat_{5/2}}_z, \ket{\rm cat_{3/2}}_z$) and the trivial $\left( \ket{1/2}+\ket{-1/2}\right) / \sqrt{2}$ state exhibit longer coherence (Fig.~\ref{fig:figure_4}e), reflecting the smaller energy difference between the $\ket{m_I}$ states involved in the superposition. $x$-oriented cat states, here prepared with the virtual-SNAP method (see Methods), exhibit longer coherence times than $z$-oriented ones, and a weaker dependence on cat size. This is consistent with the smaller population of large $|m_I|$ eigenstates in $x$-cats; most of the weight is in the $\ket{\pm1/2}$ states, which have the longest coherence time due to the first-order insensitivity to electrical noise (Supplementary Information, Section~\ref{SI_sec:T2Star}). Only dephasing (phase-flip) processes are of relevance here. The energy relaxation (bit-flip) time $T_1$ of ionised nuclear spins in silicon at low temperatures is unmeasurably long \cite{saeedi2013room,savytskyy2023electrically}. 
 
\subsection*{Outlook for quantum information processing}
 
In this work, we have demonstrated a complete experimental toolbox to exploit the 8-dimensional Hilbert space of a \Sb~nucleus. While the dimensionality may lead one to think that a $I=7/2$ nucleus is equivalent to $\log_2(2I+1)=3$ qubits, a $\ket{\rm cat_{7/2}}$ state on the nucleus is as macroscopic as a Greenberger-Horne-Zeilinger state on $2I=7$ spin-1/2 qubits and maximizes the quantum Fisher information criterion for spins \cite{frowis2018macroscopic}. Another way to appreciate this point is to observe that the $2I+1$ $\ket{I,m_I}$ eigenstates can be mapped onto the Dicke states of $2I$ spin-1/2 particles, by a process known as Dicke bootstrap \cite{kubischta2023not}, i.e. $\ket{I,m_I} \mapsto \ket{D_{I-m_I}^{2I}}$,
\begin{equation}
    \ket{D_{I-m_I}^{2I}}=\frac{1}{\sqrt{\binom{2I}{I-m_I}}}\sum_{\sigma}\sigma{\underbrace{|00\ldots0}_{I+m_I}\underbrace{11\ldots1\rangle}_{I-m_I}}, 
\end{equation}
where $\sigma$ denotes permutations over $2I$ qubits. 
A recent proposal \cite{gross2021hardware} makes use of the extreme bias in the physical noise affecting nuclear spin systems, $T_1 \gg T_2$, to encode a logical qubit in $\ket{{\rm cat}_{I}}_x$ states. The spin-cat code for $I=7/2$ can correct up to three phase flip errors (See Supplementary Information, Section\ref{sec:cat_code}). This would be surprising if one thought of the spin-7/2 as a 3-qubit equivalent, since a 3-qubit repetition code can only correct one phase flip, but understandable if considering the spin-7/2 cat state as the equivalent of a 7-qubit code. 
Here, we have demonstrated the preparation of $\ket{\rm cat_{7/2}}_x$ states that manifest the biased-noise encoding, and cat state manipulation with SU(2) rotations, showing bias-preserving logical Pauli operations.

Another proposal by Gross \cite{gross2021designing} showed that it is possible to encode in a large spin a logical qubit that satisfies the Knill-Laflamme error correction conditions \cite{knill1997theory} for first order SU(2) errors. The minimum spin size required for a code that corrects all rotation errors ($\{\hat I_x, \hat I_y,\hat I_z\}$) is $I=7/2$ \cite{gross2021designing}. 
Thus, a \Sb~nucleus can also  encode a qubit to protect information from unbiased depolarizing noise on a single spin. 
After encoding, a basis transformation from the error states to the energy eigenbasis converts rotation errors into populations of $\ket{I,m_I<I}$ eigenstates, which can be measured to detect the occurrence of the error~\cite{lim2023fault}. 
Crucially, the Gross code has SU(2) rotations as the native logical Pauli gates on the encoded qubit. Therefore, our work represents a first demonstration of such logical gates. 

Our work brings high-dimensional quantum information processing and logical qubit encoding to the realm of atomic-scale semiconductor devices. Ion-implanted donors \cite{morello2020donor} can be naturally integrated with lithographic quantum dots, which serve as a reservoir for ancillae to perform repeated rounds of error detection and correction. A dot-mediated single-spin-qudit system holds the promise to achieve beyond-break-even logical qubit lifetimes and fault-tolerant operations. Accounting for control and readout devices, and assuming medium-range coupling e.g. by electric dipole interaction \cite{tosi2017silicon} or via intermediary quantum dots \cite{wang2023jellybean}, one $^{123}$Sb qudit could occupy a footprint of order $200\times 200$~nm$^2$, affording in principle up to 25,000,000 logical qubits in a square millimetre. The technological challenges in building such a device remain formidable, but the formation of deterministic donor arrays by ion implantation is well underway \cite{jakob2023scalable}, and the technology is compatible with industry-standard metal-oxide-semiconductor (MOS) processes, which are being adapted to the development of quantum hardware \cite{zwerver2022qubits}.
 
During the preparation of this manuscript we became aware of related work, where synthetic spin-7/2 equivalents were formed using the 8-dimensional Hilbert space of a transmon qudit \cite{champion2024} and a superconducting harmonic oscillator \cite{roy2024}, and SU(2) operations and Schr\"{o}dinger cat states were similarly demonstrated. These results highlight the maturity and universality of high-dimensional quantum computing, and its readiness to underpin logical qubit platforms.

\subsection*{Methods}
\subsubsection*{Device fabrication}
The quantum processor was fabricated using standard Si MOS processes on a 900~nm thick epilayer of isotopically enriched $^{28}$Si (730~ppm residual $^{29}$Si) deposited on a natural Si handle wafer. N-type ohmic leads and p-type channels to prevent leakage currents were formed by thermal diffusion of phosphorus and boron, respectively. A central 8~nm thick high quality ultra-dry gate oxide and a surrounding 200~nm thick wet field oxide were grown in oxidation furnaces. $^{123}$Sb$^+$ ions (18~keV, $5\times10^{11}$~cm$^{-2}$) were implanted at normal incidence through a 90~nm $\times$ 100~nm implant window in a PMMA mask. A rapid thermal anneal at 1000~$^{\circ}$C for 10~s in nitrogen atmosphere was performed to repair the implantation damage and activate the donors. Surface nanoelectronics were fabricated as standard for our qubit devices using three layers of electron beam lithography and aluminium deposition. Each layer is electrically insulated from the others with native Al$_2$O$_3$, formed when aluminium is exposed to air. Finally, the sample was annealed in forming gas (400~$^{\circ}$C, 15~min, 95$\%$ N$_2$ : 5$\%$ H$_2$) to passivate interface traps.

\subsubsection*{Experimental setup}
The device was wire-bonded to a gold-plated printed circuit board and placed in a copper enclosure.
The enclosure was placed in a superconducting solenoid producing a magnetic field $B_0 = 1.384$~T (see Fig.~\ref{fig:figure_1}a for field orientation).
The board was mounted on a Bluefors BF-LD400 cryogen-free dilution refrigerator, reaching a base temperature of 18~mK.

DC bias voltages were applied to all gates using Stanford Research Systems (SRS) SIM928 voltage sources.
A room-temperature resistive combiner was used for the fast donor gates to add DC voltages to AC signals produced by two analogue output channels of the Quantum Machines OPX+, which then passed through an 80~MHz low-pass filter; all other gates passed through a 20~Hz low-pass filter. All filtering takes place at the mixing chamber plate. The wiring includes graphite-coated flexible coaxial cables to reduce triboelectric noise \cite{kalra2016vibration}.
 
Microwave pulses to induce electron spin resonance (ESR) transitions were applied to an on-chip broadband antenna \cite{dehollain2012nanoscale} using a Keysight E8267D PSG microwave signal generator.
The microwave carrier frequency remained fixed at $38.9426900$~GHz, while the output frequency was varied within a pulse sequence by mixing it with a radiofrequency (RF) signal using single-sideband modulation, i.e. by applying RF pulses to the wideband in-phase and quadrature ports of the microwave signal generator's IQ mixer. A digital output channel of the  Quantum Machines OPX+ was used to trigger the microwave signal generator. When not triggered, the carrier frequency is expected to be suppressed by 130 dB, according to the data sheet of the signal source.
The RF pulses used for single-sideband modulation were generated by two analogue output channels of the Quantum Machines OPX+. Another analogue output channel of the Quantum Machines OPX+ was used to define the generalised rotating frame and deliver multi-frequency, phase-coherent RF pulses to the microwave antenna to drive NMR transitions and apply the multi-level control Hamiltonian.
The microwave signal for ESR and RF signal for NMR were combined in a Marki Microwave DPX-1721 diplexer.
 
The SET current passed through a Femto DLPCA-200 transimpedance amplifier ($10^7$ V/A gain, 50~kHz bandwidth), followed by an SRS SIM910 JFET post-amplifier ($10^2$ V/V gain), SRS SIM965 analog filter (50 kHz cutoff low-pass Bessel filter), acquired via an analogue input port of the Quantum Machines OPX+ and then digitized.
The measurements instruments were controlled by Python code using the quantum measurement software packages QCoDeS, SilQ and QUA.
 
\subsubsection*{Measurement of dephasing in cat states}
In order to measure the dephasing of $z$-cat states in different subspaces we use a generalised Ramsey method \cite{godfrin2018generalized}.
After preparing the cat state $(\ket{m_I}+\ket{-m_I})/\sqrt{2}$ via Givens rotations, we let it evolve freely for a time $\tau$ during which it acquires a phase $\xi_{2m_I}(\tau)$.
We then undo the Givens rotations up to the $\pi/2$ pulse, which leaves the state in a superposition of $(\ket{-m_I}+e^{i\xi_{2m_I}(\tau)}\ket{-m_I+1})/\sqrt{2}$.
A final $\pi/2$ pulse with varying phase depending on $\tau$ will induce oscillations of the population in the state $\ket{-m_I}$ that decay with time.
The lifetime $T_2^\ast$ of the cat state is then defined as the time where the initial contrast of this oscillations decays to its $1/e$ point.
 
Measuring the lifetime of $x$-oriented cat states is similar to measuring the lifetime of a spin coherent state. First, applying a covariant $\pi/2$ rotation in the subspace creates the spin coherent state along $x$. By applying a virtual-SNAP gate the state is transformed into an $x-$oriented cat-state. After a free evolution time $\tau$, we undo the virtual-SNAP gate to revert the state back to a spin coherent state. A final $\pi/2$ pulse with a $\tau$-dependent phase then induces oscillations in $\langle I_z \rangle(\tau)$ from which the dephasing time $T_2^{\ast}$ is extracted.
 
When performing operations within subspaces, there is a possibility that the system will leak out of the subspace. This can be caused by imperfect state preparation, readout, or by pulse imperfections. As such leakage is easily detected by our readout process, we discard experiments where leakage has occurred.
% 
% \bibliographystyle{naturemag}
% \tocless\bibliography{bibliography.bib}
\let\oldaddcontentsline\addcontentsline% Store \addcontentsline
\renewcommand{\addcontentsline}[3]{}% Make \addcontentsline a no-op
\providecommand{\noopsort}[1]{}\providecommand{\singleletter}[1]{#1}%

\let\addcontentsline\oldaddcontentsline% Restore \addcontentsline

\subsection*{Data availability}
Source data and analysis scripts are available at \href{https://doi.org/10.5061/dryad.931zcrjtf}{https://doi.org/10.5061/dryad.931zcrjtf}. 
 
\section*{Acknowledgements}
\label{sec:acknowledgements}
We thank Jonathan A. Gross, Valerio Scarani, Jeffrey Marshall, and Jason Saied for insightful discussions. The research was funded by an Australian Research Council Discovery Project (grant no. DP210103769), the US Army Research Office (contract no. W911NF-23-1-0113) and the Australian Department of Industry, Innovation and Science (grant no. AUSMURI000002).
We acknowledge the facilities, and the scientific and technical assistance provided by the UNSW node of the Australian National Fabrication Facility (ANFF), and the Heavy Ion Accelerators (HIA) nodes at the University of Melbourne and the Australian National University. ANFF and HIA are supported by the Australian Government through the National Collaborative Research Infrastructure Strategy (NCRIS) program. Ion beam facilities employed by D.N.J. and A.M.J. were co-funded by the Australian Research Council Centre of Excellence for Quantum Computation and Communication Technology (Grant No. CE170100012).
X.Y., B.W., M.R.v.B., A.V. acknowledge support from the Sydney Quantum Academy. D.N.J. acknowledges the support of a Royal Society (UK) Wolfson Visiting Fellowship RSWVF/211016.
P.G. and B.C.S. acknowledge funding by the Natural Sciences and Engineering Research Council of Canada (NSERC), Alberta Innovates
and the Government of Alberta. N.A. is a KBR employee working under the Prime Contract No. 80ARC020D0010 with NASA Ames Research Center and is grateful for the collaborative agreement between NASA and CQC2T. The views and conclusions contained in this document are those of the authors and should not be interpreted as representing the official policies, either expressed or implied, of the Army Research Office or the U.S. Government. The U.S. Government is authorized to reproduce and distribute reprints for Government purposes notwithstanding any copyright notation herein.
Sandia National Laboratories is a multimission laboratory managed and operated by National Technology \& Engineering Solutions of Sandia, LLC, a wholly owned subsidiary of Honeywell International Inc., for the U.S. Department of Energy’s National Nuclear Security Administration under contract DE-NA0003525. This paper describes objective technical results and analysis. Any subjective views or opinions that might be expressed in the paper do not necessarily represent the views of the U.S. Department of Energy or the United States Government.

\section*{Author contributions}
X.Y., B.W., A.V., D.S., M.N., D.H., A.K., M.R.vB. and A.M. conceived and designed the experiments, with theoretical input from P.G., B.C.S, and N.A. X.Y., B.W., A.V., M.N., D.H., D.S., A.K., M.R.vB. performed and analysed the measurements. D.H. and F.E.H. fabricated the device, with A.S.D.'s supervision, on materials supplied by K.M.I.. A.M.J., D.H. and D.N.J. designed and performed the ion implantation. R.J.M., R.B.-K. and T.D.L. contributed to the data analysis. A.M., X.Y., B.W., D.H., A.V., R.B.-K., P.G., T.D.L., and N.A. wrote the manuscript, with input from all coauthors. A.M. supervised the project.

\subsection*{Competing Interests}
A.M. is an inventor on a patent related to this work, describing the use of high-spin donor nuclei as quantum information processing elements (application no. AU2019227083A1, US16/975,669, WO2019165494A1). A.S.D. is the CEO and a director of Diraq Pty Ltd.. F.E.H. and A.S.D. declare equity interest in Diraq Pty Ltd.

\newpage
\
\newpage

\beginsupplement
\title{SUPPLEMENTARY INFORMATION:\\Creation and manipulation of Schr\"{o}dinger cat states of a nuclear spin qudit in silicon}
\maketitle
\onecolumngrid
\tableofcontents
\newpage
\section{Schr\"{o}dinger cat code}
\label{sec:cat_code}
Hardware-efficient error correcting codes on a high-dimensional quantum object utilise the large Hilbert space of a single system, such as a harmonic oscillator, instead of multiple qubits to protect quantum information from dominant errors.
In particular, bosonic codes can be used to correct errors such as photon loss, photon dephasing and thermal excitations by encoding information  in the phase space of a harmonic oscillator~\cite{cai2021bosonic}.
One such code is the two-component cat code~\cite{cochrane1999macroscopically}, defined by the Schr\"{o}dinger cat states
\begin{equation}
\label{eq:bosonic-cat}
    \ket{\mathcal C_\alpha^\pm} = \frac{\ket{\alpha} \pm \ket{-\alpha}}{\sqrt{2(1\pm e^{-2|\alpha|^2})}},
\end{equation}
where $\ket{\alpha}$ is a coherent state of the bosonic mode, and $\alpha$ is an arbitrary complex number describing the displacement from the vacuum state. The 
cat state $\ket{\mathcal C_\alpha^+}$ only contains even Fock states and $\ket{\mathcal C_\alpha^-}$ only contains odd Fock states, and the encoded states are given by 
\begin{align}
    \ket{0} =& \frac{1}{\sqrt{2}} \left(\ket{\mathcal C_\alpha^+} + \ket{\mathcal C_\alpha^-}\right)
    =\ket{\alpha} + \mathcal{O}\left(\exp(-2|\alpha|^2)\right),\\
    \ket{1} =& \frac{1}{\sqrt{2}} \left(\ket{\mathcal C_\alpha^+} - \ket{\mathcal C_\alpha^-}\right)
    =\ket{-\alpha} + \mathcal{O}\left(\exp(-2|\alpha|^2)\right),
\end{align}
where the separation in the phase space between the codewords can be tuned by the amplitude $|\alpha|$ of the component coherent states. 
The bosonic cat code can suppress bit-flip errors exponentially with $|\alpha|^2$, while increasing the phase-flip error rate only linearly, and can help reduce the hardware overhead for fault tolerant quantum computation.
 
Another approach for designing hardware-efficient codes is to take advantage of the redundancy in the Hilbert space of a qudit, where the multiple levels can be used to encode information within a single finite-dimensional system.
Here, we consider an error correction code based on Schr\"{o}dinger cat states in a single spin-$I$ system, with $2I+1$ levels, for protecting information against dephasing of the spin. 
 
\subsection{Spin-cat code}
Schr\"{o}dinger cat states on a single spin are superpositions of maximally separated quasi-classical states, i.e.~spin coherent states (scs), along any axis of the spin. We consider an encoding  based on cat states along the $x$-axis
\begin{equation}
\label{eq:spincatcode}
    \ket{\rm{cat}_{\it I}^\pm}_x= \frac{1}{\sqrt2}\left(\ket{\rm{scs}_{\it I}}_{-x} \pm \ket{\rm{scs}_{\it I}}_{x}\right),
\end{equation}
where $\ket{\overline{0}}:=\ket{\rm{scs}_{\it I}}_{-x}=\ket{I,-I}_x$ and $\ket{\overline{1}}:=\ket{\rm{scs}_{\it I}}_x=\ket{I,I}_x$ are spin coherent states  oriented along the $-x$ and $+x$ axes, respectively, that define the logical encoding. 
The cat state $\ket{\rm{cat}_{\it I}^+}_{x}$ only contains even parity levels and $\ket{\rm{cat}_{\it I}^-}_{x}$ only contains odd parity levels, similar to bosonic cat states~\eqref{eq:bosonic-cat}. 
In the basis $\{\ket{I,m_I}_z\}$ of the eigenstates of $\hat I_z$, corresponding to the physical quantization axis set by the nuclear Zeeman energy, the encoded states are given by
\begin{align}
\label{eq:catcodewords}
     \ket{\overline0}=& \sum_{m_I=-I}^I d_{m_I,-I}^I\left(\frac{\pi}{2}\right)\ket{I,m_I}_z,\\
     \ket{\overline1}=& \sum_{m_I=-I}^Id_{m_I,I}^I\left(\frac{\pi}{2}\right)\ket{I,m_I}_z,
\end{align}
where $d_{m',m}^I\left(\beta=\pi/2\right)$ is the Wigner d-function~\cite{biedenharn1984angular}. It is noteworthy that the codewords $\ket{\overline{0}}$ and $\ket{\overline{1}}$ for the spin-cat code are perfectly orthonormal to each other, unlike the bosonic cat code, which has $\mathcal{O}\left(\exp(-2|\alpha|^2)\right)$ overlap between the encoded states. 
\subsection{Error correction}
The physically dominant noise in spin systems arises from local fluctuations in the surrounding magnetic and electric fields, resulting in decoherence of the spins \cite{muhonen2014storing}. 
In particular, for a nuclear spin, errors along the physical quantization axis, i.e.~the $z$ axis, dominate over the errors along other axes, leading to a biased-noise system that could be used to reduce the overhead in fault tolerant quantum computation, similar to the bosonic case.
The Schr\"{o}dinger spin-cat code on a single high-spin nucleus can be used to correct biased errors that cause dephasing, described by the angular momentum operator $\hat I_z$.  After an $\hat I_z$ error, the nuclear spin state is given by
\begin{align}
    \hat I_z\ket{\overline{0}} =& \hat U^\dagger \left(\hat U\hat I_z \hat U^\dagger\right) \hat U\ket{I,-I}_x = \hat U^\dagger\hat I_x \ket{I,-I}_z = \hat U^\dagger \ket{I,-I+1}_z = c_1\ket{I,-I+1}_x, \\
    \hat I_z\ket{\overline{1}} =& \hat U^\dagger \left(\hat U\hat I_z \hat U^\dagger\right) \hat U\ket{I,I}_x = \hat U^\dagger\hat I_x \ket{I,I}_z = \hat U^\dagger \ket{I,I-1}_z = c_1\ket{I,I-1}_x, 
\end{align}
where $c_1$ is a complex number, $\hat U = e^{-i\frac\pi 2 \hat I_y}$ and $\hat U \hat I_z \hat U^\dagger = \hat I_x$. Similarly, after an $\hat I_z^2$ error, the encoded state is given by
\begin{align}
    \hat I_z^2 \ket{\overline{0}}=& \hat I_z^2\ket{I,-I}_x=c_2\ket{I,-I}_x+c_3\ket{I,-I+2}_x\\
    \hat I_z^2 \ket{\overline{1}}=& \hat I_z^2\ket{I,I}_x=c_2\ket{I,I}_x+c_3\ket{I,I-2}_x,
\end{align}
where $c_2$ and $c_3$ are complex numbers.  For spin $I\geq{5}/{2}$, the codewords are orthogonal to each other for both linear and quadratic forms of the error $\hat I_z$, which can be corrected using the spin-cat code. In fact, a spin-$7/2$ cat code can also correct for  $\hat I_z^3$ type errors, so, the set of correctable errors is spanned by $\{\mathbb{I}, \hat I_z, \hat I_z^2, \hat I_z^3\}$. 
The Schr\"{o}dinger spin-cat code can exponentially suppress bit-flip errors as the spin $I$ increases, with only a linear increase in phase-flip errors on the encoded qubit, similar to the bosonic-cat code. 
 
\subsection{Logical gates}
Logical operations that process information encoded in a spin-cat qubit should preserve the noise bias, i.e.~
\begin{equation}
    \overline U \hat E_i \overline U^\dagger \propto \hat E_i,
\end{equation}
where $\hat E_i$  denotes the error operator and $\overline U$ is a logical operation on the encoded qubit. Covariant $\pi$-rotations $\overline U = e^{i\pi \hat I_\alpha }$ about any axis $\alpha \in \{x,y,z\}$ satisfy this condition, and can be used to implement logical Pauli operations on the encoded qubit. 
In particular, the logical $\overline X$ gate can be implemented by a covariant $\pi$-rotation about the $z$ axis, which applies the transformation
\begin{align}
    e^{-i\pi \hat I_z}\ket{\overline{0}} =& \sum_{m_I=-I}^I(-1)^{m_I}d_{m_I,-I}^I\left(\frac{\pi}{2}\right) \ket{I,m_I}_z\equiv \ket{\overline{1}},\\
    e^{-i\pi \hat I_z}\ket{\overline1}=& \sum_{m_I=-I}^{I}(-1)^{m_I}d_{m_I,I}^I\left(\frac{\pi}{2}\right)\ket{I,m_I}_z\equiv \ket{\overline{0}},
\end{align}
where we used Eq.~\eqref{eq:catcodewords} along with the relation $e^{-i\pi \hat I_z}\ket{I,m_I}_z = e^{-i\pi m_I}\ket{I,m_I}_z = (-1)^{m_I}\ket{I,m_I}_z$. 
Here, one can see that  $e^{-i\pi \hat I_z}$ is equivalent to the logical Pauli gate $\overline{X}$. 
Similarly, the logical $\overline Z$ gate can be implemented using a covariant $\pi$ rotation about the $x$-axis leading to the transformation
\begin{align}
    e^{-i\pi \hat I_x}\ket{\overline{0}} =&  e^{-i\pi \hat I_x}\ket{I,-I}_x = (-1)^{-I} \ket{I,-I}_x \equiv \ket{\overline{0}},\\
    e^{-i\pi \hat I_x}\ket{\overline{1}} =&  e^{-i\pi \hat I_x}\ket{I,I}_x = (-1)^{I} \ket{I,I}_x\equiv-\ket{\overline{1}},
\end{align}
where we used Eq.~\eqref{eq:catcodewords} along with the relation $e^{-i\pi \hat I_x}\ket{I,m_I}_x = e^{-i\pi m_I}\ket{I,m_I}_x$. It should be noted that the last equality only holds for half-odd integer spins. 
Thus, covariant rotations of the form $\overline{U}=e^{i\pi \hat I_\alpha }$ can be used to implement logical Pauli operations on the encoded state in bias-preserving way, allowing the Schr\"{o}dinger spin-cat code to be concatenated with multi-qubit codes to achieve high-threshold fault tolerant quantum computation.

\section{Spin Wigner function}
\label{representations}
\subsection{Definition}
The Wigner function is a widely used tool for representing the joint quasi-probability distribution of position and momentum variables, denoted as $\{\hat{x},\hat{p}\}$, on a planar phase space. Because $\hat{x}$ and $\hat{p}$ do not commute in quantum mechanics, the value of the Wigner function can be negative. Negativity is commonly regarded as a signature of nonclassicality.
Here, we compute and use the Wigner function \cite{ferrie2011quasi} for a single high-spin nuclear system, which can be described by spin operators ${\hat{I}_x,\hat{I}_y, \hat{I}_z}$ with a fixed length
$\mathrm{I}=\vert{\hat{I}}\vert=\sqrt{\langle{\hat{I}}_x\rangle^2 + \langle {\hat{I}}_y \rangle^2 +\langle{\hat{I}}_z\rangle^2}$
and commuting relations for all cyclic permutations $[\hat{I}_\alpha,\hat{I}_\beta] = i\hbar \hat{I}_\gamma$, $\alpha,\beta,\gamma =\{x,y,z\}$.
These spin operators can be used to define the Wigner function, denoted as $W_{\rho}(\theta, \varphi)$, on a spherical phase space, where $\rho$ is the density matrix, $\theta$ is the polar angle measured along the +z~axis, and $\varphi$ is the azimuth angle \cite{klimov2017generalized}.
%\comm{citation Phys. A: Math. Theor. 50, 323001 (2017)}. 
More specifically, $W_{\rho}(\theta, \varphi)$ is defined as:
\begin{equation} 
\label{spin Wigner 1}
    W_{\rho}(\theta, \varphi) = \sqrt{\frac{2}{\pi}}\ 
    \sum_{k=0}^{2I}\sum_{q=-k}^{k} Y_{kq}(\theta,\varphi)\rho_{kq} \
\end{equation}
where $Y_{kq}$ are the spherical harmonics and $\rho_{kq}$ is an element of the density matrix decomposed in the spherical harmonic basis \cite{Agarwal1981}. This equation shows that $W$ is entirely determined by the density matrix $\rho$ and can therefore be reconstructed using quantum state tomography (see~\ref{Density matrix reconstruction}).
 
\subsection{Visualisation}
\begin{figure}[ht!]
    \centering
    \includegraphics[width=13cm]{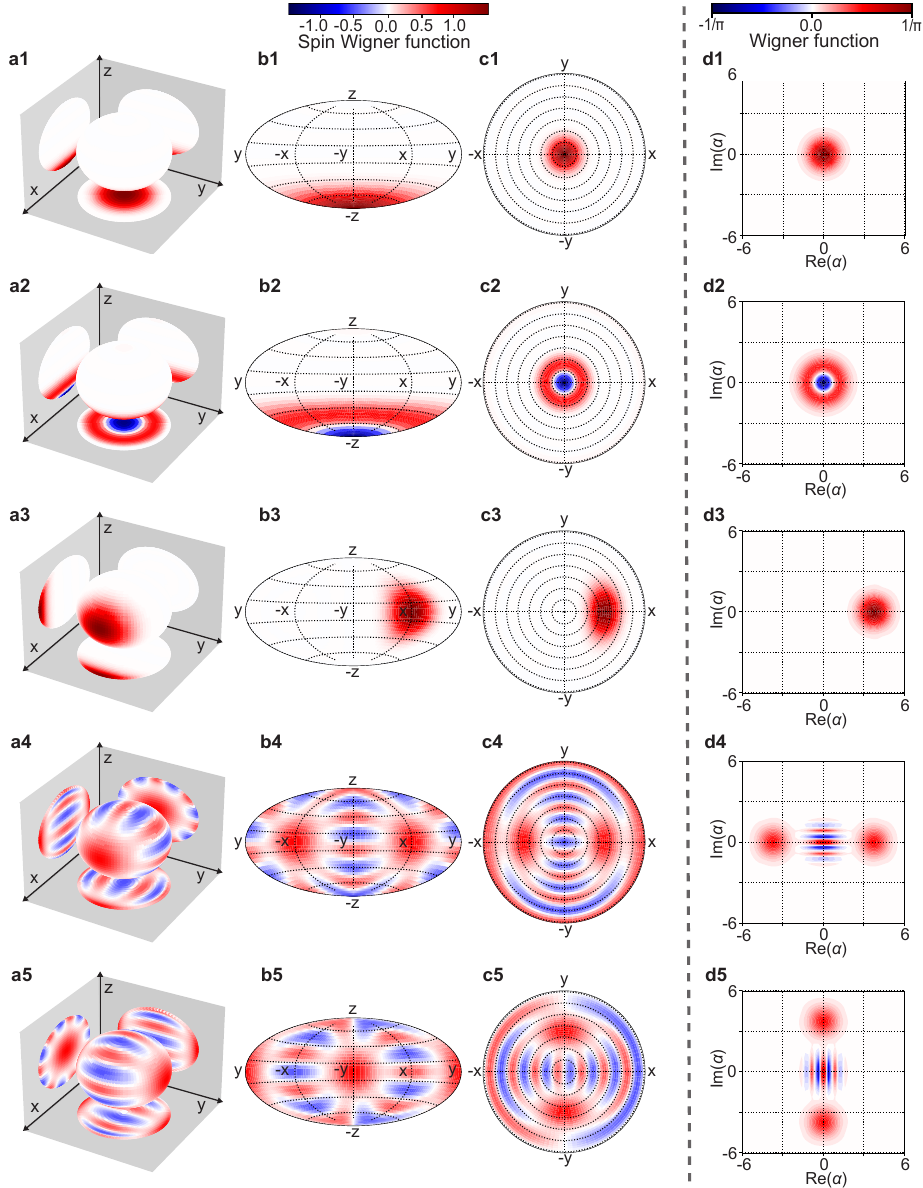}
    \caption{\textbf{Comparison of Wigner functions in high-spin states and bosonic states in quantum optics}. 
    \textbf{a}, 3D spherical spin Wigner functions in an 8-dimensional Hilbert space. The grey planes in the Wigner plots show the mirrored reflections of the sphere. The projection plots
    \textbf{b}, 2D Hammer projections of the spherical Wigner functions.
    \textbf{c}, 2D Polar projections of the spherical Wigner functions, observed from the south pole (the north pole becomes the outer circle).
    \textbf{d}, Wigner functions for a bosonic mode, where $\alpha$ is an arbitrary complex number, describing the displacement from the vacuum state. The following states are represented: 
    \textbf{a1-c1}, $\ket{-7/2}$; 
    \textbf{d1}, vacuum state $\ket{0}$,
    \textbf{a2-c2}, $\ket{-5/2}$, containing one spin excitation from the $\ket{-7/2}$ state;
    \textbf{d2}, Fock state $\ket{1}$, i.e. a bosonic mode containing one photon;
    \textbf{a3-c3}, $\ket{\text{scs}_{7/2}}_x$, a spin coherent state pointing maximally along $x;$ 
    \textbf{d3}, displaced coherent state $(\hat{\mathcal{D}}(\sqrt{7})\ket{0}$. In rows 1,2 and 3, note the similarity between spin and bosonic states, especially evident in the polar projections.
    \textbf{a4-c4}, $x$-oriented Schr\"{o}dinger cat state $\ket{\rm cat_{7/2}}_x = \left( \ket{7/2}_x+\ket{-7/2}_x\right) / \sqrt{2}$;
    \textbf{d4}, $x$-oriented bosonic cat state $\frac{1}{\sqrt{2}}(\hat{\mathcal{D}}(\sqrt{7}i)\ket{0}+\hat{\mathcal{D}}(-\sqrt{7}i)\ket{0})$,
    \textbf{a5-c5}, $y$-oriented Schr\"{o}dinger cat state $\ket{\rm cat_{7/2}}_y = \left( \ket{7/2}_y+\ket{-7/2}_y\right) / \sqrt{2}$,
    \textbf{d5}, $p$-oriented bosonic cat state $\frac{1}{\sqrt{2}}(\hat{\mathcal{D}}(\sqrt{7})\ket{0}+\hat{\mathcal{D}}(-\sqrt{7})\ket{0})$.
    }
    \label{fig:WignerPlot}
\end{figure}

The Wigner function serves as a convenient tool for visualising states in high-dimensional systems, particularly for highlighting highly non-classical states that exhibit distinct fringe patterns. Despite the 
%discrete and 
%%% RBK says "I can't add Overleaf comments for some reason, but I removed "discrete" because quantum optical states are also "discrete" (the number basis is discrete), just not finite."
finite-dimensional nature of high-spin states, they bear a strong resemblance to quantum optical states in 2D continuous phase space. In Fig.~\ref{fig:WignerPlot} we provide examples of 3D spherical Wigner functions of high-spin states, complemented by their corresponding planar representations utilising Hammer and polar projections. In addition, we provide 2D Wigner functions representing bosonic states in quantum optics to provide a side-by-side comparison and highlight a few compelling analogies with high-spin states.
In particular, there is a clear resemblance between spin coherent states and bosonic coherent states, or between spin projection eigenstates $\ket{m_I}$ and bosonic Fock states $\ket{N}$. This analogy is particularly evident in the polar projection for spin states. For Schr\"{o}dinger cat states, spin and bosonic states exhibit similar interference patterns, which can be oriented along different directions -- $x$ and $z$ for spins, $x$ (or Re$(\alpha)$) and $p$ (or Im$(\alpha)$) for bosons. Here, the analogy is most evident when comparing the spin Hammer projection to the phase-space representation of bosonic modes.
 
However, the analogy is not all-encompassing. For example, the action of a bosonic displacement operator, $\hat{\mathcal{D}}$, is closely analogous to the effect of a covariant SU(2) rotation (CR) on a spin coherent state, but this optical analogy does not hold when considering the fact that a covariant $\pi/2$ spin rotation turns $\ket{\rm cat_{\it I}}_x$ to $\ket{\rm cat_{\it I}}_y$. In the bosonic case, the rotation would require a phase shift instead of a displacement.
 
\newpage
 
\subsection{Phase shifts and fringe patterns in the Wigner function}
The fringe pattern in the Wigner function captures the phase information contained in a quantum state. Consider the two-component spin cat states $\ket{\rm{cat_{7/2}}}_{\mathrm{n}}$, which take the form
 \begin{equation}
    \ket{\rm{cat}_{7/2}}_{\mathbf{n}} = \frac{1}{\sqrt{2}}
    \left( \ket{\rm{scs}_{7/2}}_{\mathbf{n}} + e^{i\xi_7}\ket{\rm{scs}_{7/2}}_{-\mathbf{n}} \right),
\end{equation}
\noindent
where $\ket{\rm{scs}_{7/2}}_{\mathbf{n}}$ and $\ket{\rm{scs_{7/2}}}_{-\mathbf{n}}$ are two anti-parallel spin coherent states, oriented along the unit vectors $\mathbf{n}$ and $-\mathbf{n}$ in the spin-7/2 Bloch sphere. The dependence on the phase factor $e^{i\xi_7}$ indicates that a cat state is \emph{not} uniquely defined by the position of its `heads'. We further illustrate this point by inspecting the Wigner functions of cat states of along the $z$ and $x$ axes, denoted as $\ket{\rm cat_{7/2}}_z$ and $\ket{\rm cat_{7/2}}_x$ respectively.
 
\begin{figure}[ht!]
    \centering
    \includegraphics[width=14cm]{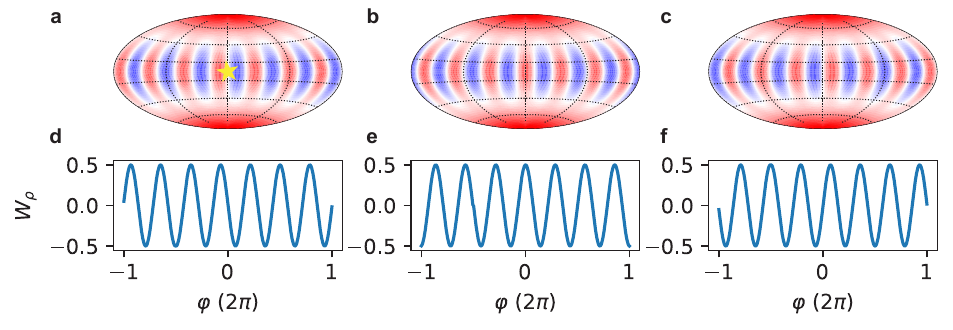}
    \caption{\textbf{Phase dependence of the Wigner function of a $z$-oriented spin-cat state}. 
    Hammer projection of the Wigner function of $\ket{\rm cat_{7/2}}_z = \left( \ket{7/2}+e^{i\xi_7}\ket{-7/2}\right) / \sqrt{2}$ for \textbf{a} $\xi_7 = 0$, \textbf{b} $\xi_7 = \pi/2$, and \textbf{c} $\xi_7 = \pi$.
    \textbf{d-f} Equatorial line-cuts of the Wigner functions in \textbf{a-c}.% $W_{\rho}(\pi/2, \varphi)$. Here $\rho$ is the density matrix of the $\ket{\rm cat_{7/2}}_z$.%Please note, that even though the three cat states shown in panels \textbf{a}, \textbf{b}, and \textbf{c} display high-degree visual similarity, they differ significantly in the phase of the equatorial oscillation, which can be seen in the line-cuts.
    }
    \label{fig:z_cat}
\end{figure}

In Fig.~\ref{fig:z_cat} we plot the Wigner functions $W_{\rho}(\theta, \varphi)$ of three $z$-cats, distinguished by their different values of $\xi_7 = {0, \pi/2, \pi}$. Under each Hammer projection of the Wigner plot, we draw a line-cut around the equator. This figure highlights that all $\ket{\rm cat_{7/2}}_z$ states always display 7 periods of oscillations around the equator, and the phase $\xi_7$ determines the phase of such oscillations. Therefore, even though all these states have their `heads' located at $+z$ and $-z$, they can be distinguished by the phase of the oscillations of the Wigner function around the equator.

\begin{figure}[ht!]
    \centering
    \includegraphics[width=6cm]{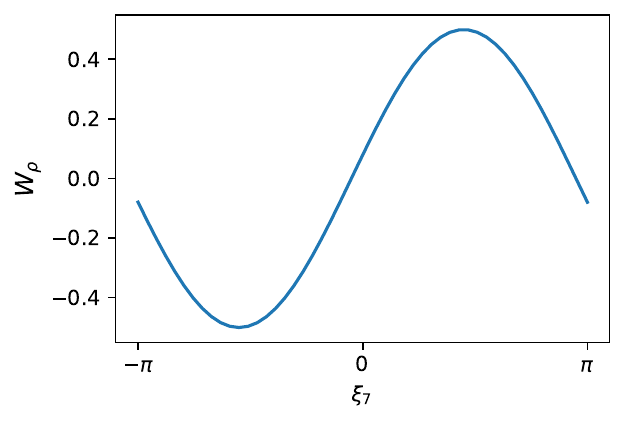}
    \caption{\textbf{Wigner function of a $z$-cat at $(\pi/2, -\pi/2)$ as a function of phase $\xi_7$.}}
    % A complete $2\pi$ sweeping of $\xi_7$ for $\ket{\rm cat_{7/2}}_z = \left( \ket{7/2}+e^{i\xi_7}\ket{-7/2}\right) / \sqrt{2}$
    \label{fig:cat_z_phase}
\end{figure}

In Fig.~\ref{fig:cat_z_phase} we plot the value of the Wigner function of $\ket{\rm cat_{7/2}}_z = \left( \ket{7/2}+e^{i\xi_7}\ket{-7/2}\right) / \sqrt{2}$ at a fixed location on the Bloch sphere, namely $(\theta = \pi/2, \varphi = -\pi/2)$, (marked as a yellow star in Fig.~\ref{fig:z_cat}a), as a function of $\xi_7 \in [-\pi,\pi]$. Here we note that a full $2\pi$ sweep of the phase $\xi_7$  results in only one period of oscillation of $W_{\rho}(\pi/2, -\pi/2)$, despite the presence of 7 oscillations along the equatorial axis $\varphi$ in Fig.~\ref{fig:z_cat}a. In other words, it only takes 1/7 of a rotation of the Wigner function to cover the whole range of phases $\xi_7$. 
 
\begin{figure}[ht!]
    \centering
    \includegraphics[width=14cm]{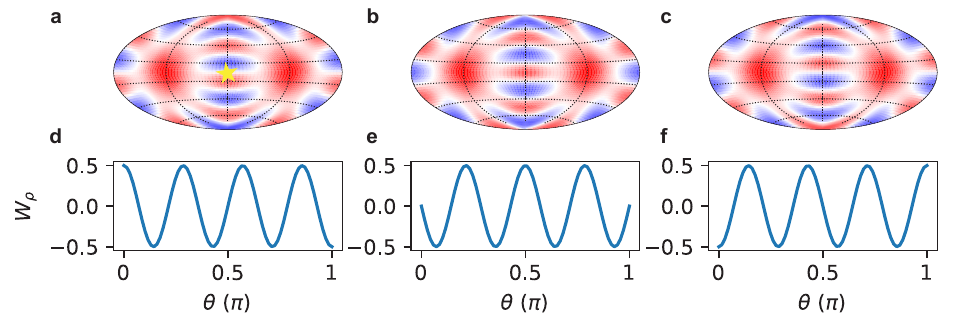}
    \caption{\textbf{Phase dependence of the Wigner function of a spin-cat state along x axis}. 
    \textbf{a-c} show the plots of the Wigner function of $\ket{\rm cat_{7/2}}_x = \left( \ket{7/2}_x+e^{i\xi_7}\ket{-7/2}_x\right) / \sqrt{2}$ for different phases $\xi_7$. \textbf{a}, $\xi_7 = 0$,\textbf{b}, $\xi_7 = \pi/2$, and \textbf{c}, $\xi_7 = \pi$.
    \textbf{d-f} show the corresponding polar line-cuts of the Wigner function $W{\rho}(\theta, -\pi/2)$ highlighting the oscillation in the Wigner function along the polar line.
    }
    \label{fig:x_cat}
\end{figure}

\begin{figure}[ht!]
    \centering
    \includegraphics[width=6cm]{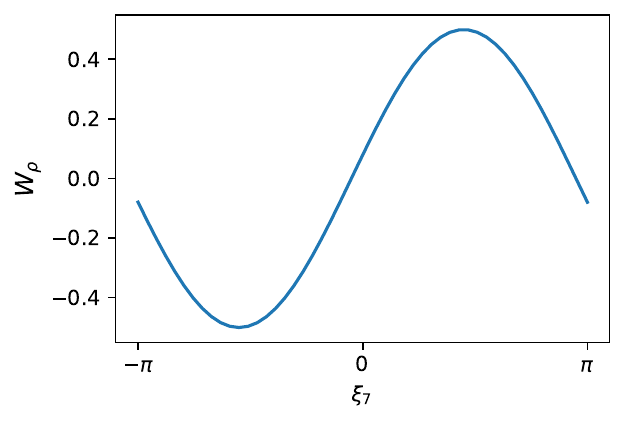}
    \caption{\textbf{Wigner function of an $x$-cat at $(\pi/2, -\pi/2)$ as a function of phase $\xi_7$.} 
    }
    \label{fig:cat_x_phase}
\end{figure}

Next we consider the cat states along the $x$ direction, $\ket{\rm cat_{7/2}}_x = \left( \ket{\rm{scs}_{7/2}}_{x}+e^{i\xi_7}\ket{\rm{scs}_{7/2}}_{-x}\right) / \sqrt{2}$, where $\ket{\rm{scs}_{7/2}}_{\pm x}$ represent spin coherent states along the $\pm x$ direction. %, which can be generated by $\pi/2$ covariant SU(2) rotations around the $y$ axis of the eigenstates $\ket{7/2}$ and $\ket{-7/2}$ respectively. 
We illustrate three examples of $\ket{\rm cat_{7/2}}_x$ states for different phases, $\xi_7 = 0, \pi/2, \pi$. In the notation used in Section~\ref{sec:cat_code}B, $\xi_7 = 0$ yields the $\ket{\rm cat_{7/2}^+}_x$ state, and $\xi_7 = \pi$ yields the $\ket{\rm cat_{7/2}^-}_x$. As shown in in Fig.~\ref{fig:x_cat}, the `heads' of the cat remain fixed regardless of the phase $\xi_7$, whereas the fringe patterns along the polar axis $\theta$ vary with $\xi_7$. Line-cuts along the meridian at fixed longitude $\varphi = -\pi/2$ display oscillations in the Wigner function whose phase depends on the chosen value of $\xi_7$, just like in the $z$-cat case. Since we plot $\theta$ from 0 to $\pi$, only $3.5$ oscillation periods are seen instead of $7$. Again, fixing the point $(\theta = \pi/2, \varphi = -\pi/2)$ (yellow star in Fig.~\ref{fig:x_cat}a) and sweeping the phase $\xi_7 \in [-\pi,\pi]$ in the state $\ket{\mathrm{cat}_{7/2}}_x$ results in a single-period oscillation for $W_{\rho}(\pi/2, -\pi/2)$ (Fig.~\ref{fig:cat_x_phase}).
 
\begin{figure}[ht!]
    \centering
    \includegraphics[width=10cm]{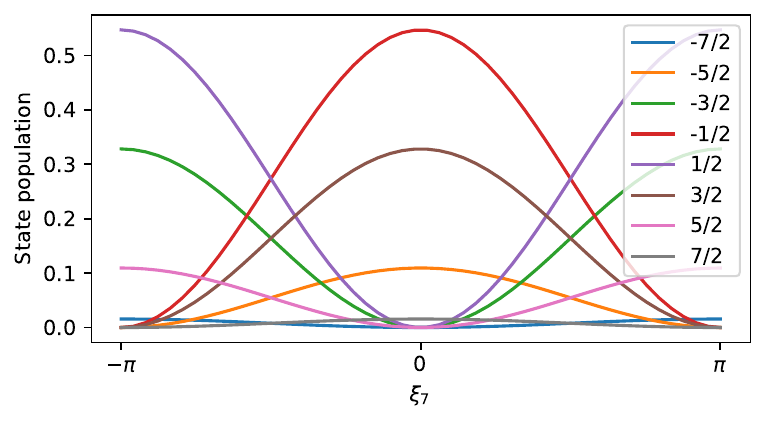}
    \caption{\textbf{Phase-dependent state populations for an $x$-cat.} The population of the $\ket{m_I}$ states is plotted as a function of the phase $\xi_7$ for a $\ket{\mathrm{cat}_{7/2}}_x$ state.}
    \label{fig:cat_x_population}
\end{figure}

There is, however, a crucial difference between $x$-cats and $z$-cats. Whereas in the $z$ case the phase $\xi_7$ only determines the phase of the Wigner function oscillations, in the $x$ case it also \emph{changes the populations} of the eigenstates $\ket{m_I}$ (Fig.~\ref{fig:cat_x_population}) and, as consequence, the parity of the state. Below we show explicitly the populations of three representative states, for $\xi_7 = 0$, $\xi_7 = \pi/2$, and $\xi_7 = \pi$, resulting in even, zero, and odd parities.
 
\begin{subequations}
\begin{minipage}{0.2\linewidth}
\begin{equation}
  \ket{\mathrm{cat}_{7/2}, \xi_7 = 0}_x = 
  \left(
  \begin{array}{c}
    0.125 \\
    0 \\
    0.573 \\
    0 \\
    0.740 \\
    0 \\
    0.331 \\
    0 \\
  \end{array}
  \right)
\end{equation}
\end{minipage}
\hspace{\fill}
\begin{minipage}{0.4\linewidth}
\begin{equation}
  \ket{\mathrm{cat}_{7/2}, \xi_7 = \pi/2}_x = 
  \left(
  \begin{array}{c}
    -0.088 \times e^{-i\frac{3\pi}{4}} \\
    0.234 \times e^{-i\frac{\pi}{4}} \\
    -0.405 \times e^{-i\frac{3\pi}{4}} \\
    0.523 \times e^{-i\frac{\pi}{4}} \\
    -0.523 \times e^{-i\frac{3\pi}{4}} \\
    0.405 \times e^{-i\frac{\pi}{4}} \\
    -0.234 \times e^{-i\frac{3\pi}{4}} \\
    0.088 \times e^{-i\frac{\pi}{4}} \\
  \end{array}
  \right)
\end{equation}
\end{minipage}
\hspace{\fill}
\begin{minipage}{0.3\linewidth}
\begin{equation}
  \ket{\mathrm{cat}_{7/2}, \xi_7 = \pi}_x = 
  \left(
  \begin{array}{c}
    0 \\
    0.331 \\
    0 \\
    0.740 \\
    0 \\
    0.573 \\
    0 \\
    0.125 \\
  \end{array}
  \right)
\end{equation}
\end{minipage}
\end{subequations}
 
In particular, when $\xi_7 = \pm \pi/2 $, the cat states along $x$ direction exhibit the same state population as the spin coherent state along the equator. Within the generalised rotating frame, a virtual-SNAP gate can only create cat states with identical populations as a spin coherent state along the equator, as detailed in Section~\ref{SNAP}.
 
\section{Covariant SU(2) rotation theory}
\label{sec:GlobalRotationTheory}
 
\subsection{Generalised rotating frame}
 
The dynamics of any system are relative to a specific frame of reference \cite{giacomini2019quantum}. Quantum systems governed by their respective Hamiltonians are no exception. In the laboratory frame, the relative phase of a qubit accumulates at the resonance frequency determined by the energy difference between its two levels. However, one can define a rotating frame that counters this phase evolution, effectively altering the qubit's dynamics by transitioning to a different frame. Typically, the reference frequency of the rotating frame is selected to be equal to the qubit's resonance frequency, facilitating simpler dynamics within the rotating frame as the accumulating phase of the qubit is effectively cancelled out.

In higher-dimensional spin systems, the presence of multiple resonance frequencies means that no single reference frequency can cancel all relative phases between multiple energy levels. Nonetheless, the concept can be expanded to encompass a generalised rotating frame (GRF), which utilises multiple reference frequencies to account for different energy splittings between pairs of adjacent energy levels \cite{leuenberger2003grover}. Here the GRF is tailored for an $n$-level non-degenerate quantum system with $n-1$ neighboring transitions, each of which is characterized by resonance frequencies $f^0_{i}$.  
We apply the generalised rotating frame transformation, which converts the Hamiltonian in the laboratory frame denoted as $\hat{\mathcal{H}}_{\rm{lab}}$ into the Hamiltonian in the generalised rotating frame, $\hat{\mathcal{H}}_{\rm{GRF}}$, through the unitary transformation:
  \begin{equation}
    \hat{\mathcal{H}}_{\rm{GRF}}^{} = \hat{\mathcal{U}}_{\rm{GRF}}^\dagger
    \hat{\mathcal{H}}_{\rm{lab}}^{} \hat{\mathcal{U}}_{\rm{GRF}}^{} -
      i \hbar \hat{\mathcal{U}}_{\rm{GRF}} ^\dagger\frac{\partial \hat{\mathcal{U}}_{\rm{GRF}}}{\partial t} ,
      \label{HDS-eq:unitary transformation}
  \end{equation}
where the unitary GRF operator $\hat{\mathcal{U}}_{\rm{GRF}}$ is defined as
  \begin{equation}
      \hat{\mathcal{U}}_{\rm{GRF}} = 
  \left(
  \begin{array}{cccc}
   e^{-i 2\pi f_1^\mathrm{ref} t} & 0 & 0 & 0 \\
   0 & e^{-i 2\pi f_2^\mathrm{ref} t} & \cdots & 0 \\
   0 & \vdots & \ddots & 0 \\
   0 & 0 & 0 & e^{-i 2\pi f_n^\mathrm{ref} t}
  \end{array}
  \right).
  \end{equation}
Within the GRF framework, $n-1$ distinct reference frequencies ($f_{i}^\text{ref}$) are required to fully counteract the $n-1$ resonance frequencies ($f^0_{i}$). Correspondingly, there are $n-1$ frequency detunings, denoted as $\delta_{i} = f^\text{ref}_{i} - f^0_{i}$.  If the Hamiltonian is time-independent in the laboratory frame -- in our case, in the absence of NMR driving terms -- the GRF with zero detunings cancels it out completely, so that any state appears static within that frame. 
 
\subsection{Multi-frequency driving pulse}
\label{multi-tone drive}
We now elucidate the system dynamics within the GRF framework when time-dependent driving terms are introduced. We confine our analysis to the ionised $^{123}$Sb nuclear spin, but note that this concept is broadly applicable to various high-level systems.
 
In the laboratory frame, the time-dependent Hamiltonian of a high-spin system is expressed as follows:
\begin{equation}
  \hat{\mathcal{H}}_{\rm{lab}}(t) = \hat{\mathcal{H}}_{D^+}
  -\gamma_\mathrm{n} \hat{I}_x \sum_{k=1}^{2I}\cos(2\pi f_k t + \phi_k)B_{1,k}(t),
  \label{HDS-eq:grf general Hamiltonian}
\end{equation}
where $\hat{\mathcal{H}}_{D^+}$ (Eq.~(1) in the main text) is the static term of the ionised donor Hamiltonian, set by the Zeeman energy and the quadrupole interaction. In time-dependent term, $f_k$, $\phi_k$, and $B_{1,lk}(t)$ represent the frequency, phase, and amplitude of each tone in the driving pulse, respectively, and $\hat{I}_x$ is the high spin operator along the $x$ direction. For the spin-7/2 system, 
\begin{align}
\hspace{-2cm}
\hat{I}_x^{7/2} = \frac{1}{2}
  \left(
  \begin{array}{cccccccc}
  0 & \sqrt{7} & 0 & 0 & 0 & 0 & 0 & 0 \\
  \sqrt{7} & 0 & \sqrt{12} & 0 & 0 & 0 & 0 & 0 \\
  0 & \sqrt{12} & 0 & \sqrt{15} & 0 & 0 & 0 & 0 \\
  0 & 0 & \sqrt{15} & 0 & \sqrt{16} & 0 & 0 & 0 \\
  0 & 0 & 0 & \sqrt{16} & 0 & \sqrt{15} & 0 & 0 \\
  0 & 0 & 0 & 0 & \sqrt{15} & 0 & \sqrt{12} & 0 \\
  0 & 0 & 0 & 0 & 0 & \sqrt{12} & 0 & \sqrt{7} \\
  0 & 0 & 0 & 0 & 0 & 0 & \sqrt{7} & 0 \\
  \end{array}
  \right).
  \label{Ix_spin_operator}
\end{align}
 
Converting Eq.~\eqref{HDS-eq:grf general Hamiltonian} into a time-independent Hamiltonian in the generalised rotating frame relies on two assumptions:
\begin{itemize}
    \item Rotating wave approximation (RWA). The driving field in Eq.~\eqref{HDS-eq:grf general Hamiltonian} can be decomposed into $2\times2I$ circular rotating fields with amplitude $B_{1,k}$ and with opposite rotating directions, represented by the frequencies $f_{k}$ and $-f_{k}$. The RWA results in the reduction of $B_{1,k}$ by a factor of two by eliminating the counter-rotating terms with frequencies $-f_{k}$. The RWA is well obeyed in the present experiment, since the driving strength is intentionally set in the sub-kilohertz regime, making it $\sim 10,000\times$ smaller than the NMR frequencies, $\approx 7.7$~MHz.
    \item Each tone $k$ drives only a single transition, leaving all other states unaffected. This condition holds true when the driving strengths are much lower than the separations between resonance frequencies. With the increase of the driving strength, this assumption breaks down and the resulting Hamiltonian is detailed in Section~\ref{sec:GlobalRotationTheory}C. In our experiment, the frequency separation between each transition is determined by the quadrupole interaction (28~kHz), while the driving strength is set in the sub-kilohertz regime. 
\end{itemize}
 
The resulting generalised rotating frame Hamiltonian has the form:
{\small
%\begin{align}
%\hspace{-1.5cm}
\begin{multline}
\hat{\mathcal{H}}_{\mathrm{GRF}}= 
\\-\frac{\gamma_{\mathrm{n}}}{4}
  \left(
  \begin{array}{cccccccc}
  0 & \sqrt{7}B_{1,1}e^{i\phi_1} & 0 & 0 & 0 & 0 & 0 & 0 \\
  \sqrt{7}B_{1,1}e^{-i\phi_1} & \delta_1 & \sqrt{12}B_{1,2}e^{i\phi_2} & 0 & 0 & 0 & 0 & 0 \\
  0 & \sqrt{12}B_{1,2}e^{-i\phi_2} & \delta_2 & \sqrt{15}B_{1,3}e^{i\phi_3} & 0 & 0 & 0 & 0 \\
  0 & 0 & \sqrt{15}B_{1,3}e^{-i\phi_3} & \delta_3 & \sqrt{16}B_{1,4}e^{i\phi_4} & 0 & 0 & 0 \\
  0 & 0 & 0 & \sqrt{16}B_{1,4}e^{-i\phi_4} & \delta_4 & \sqrt{15}B_{1,5}e^{i\phi_5} & 0 & 0 \\
  0 & 0 & 0 & 0 & \sqrt{15}B_{1,5}e^{-i\phi_5} & \delta_5 & \sqrt{12}B_{1,6}e^{i\phi_6} & 0 \\
  0 & 0 & 0 & 0 & 0 & \sqrt{12}B_{1,6}e^{-i\phi_6} & \delta_6 & \sqrt{7}B_{1,7}e^{i\phi_7} \\
  0 & 0 & 0 & 0 & 0 & 0 & \sqrt{7}B_{1,7}e^{-i\phi_7} & \delta_7 \\
  \end{array}
  \right)
  \label{GRF_H}
\end{multline}
}
 
In our implementation, a Quantum Machines OPX+, an FPGA-based signal generator, manages the amplitude $B_{1,k}$ of each tone, enabling individual tuning of the coupling strengths between energy levels. In addition, the signal generator creates the generalised rotating frame by defining 7 internal `virtual clocks', i.e. 7 software-defined oscillators with a well-defined phase relationship with each other. With 14 independent parameters, $B_{1,k}$ and $\phi_{k}$, this setup enables the covariant SU(2) rotations of arbitrary angle $\Theta$ around an arbitrary equatorial axis $\varphi$ within the GRF: 
\begin{equation}
    R_{\Theta}(\varphi) = e^{i\Theta(\hat{I}_x \cos{\varphi} + \hat{I}_y\sin{\varphi})}.
\end{equation}
 
Furthermore, manipulating the reference phases at the software level facilitates the virtual-SNAP gate (Supplementary Information, Section ~\ref{SNAP}). With these control capabilities, we can implement arbitrary axis covariant SU(2) rotations, subspace rotations, and virtual-SNAP gate within the generalised rotating frame. 
 
\subsection{Power dependence of covariant SU(2) rotations}
\label{sec:GRvsPower}
\begin{figure}[ht!]
    \centering
    \includegraphics{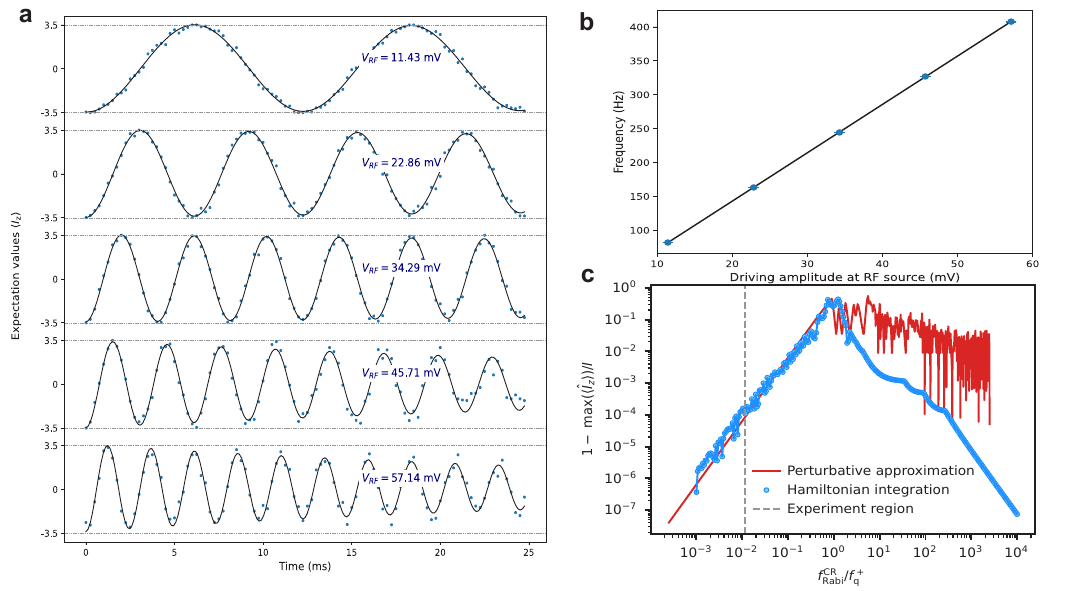}
    
    \caption{
    \textbf{Dependence of covariant SU(2) rotations on driving amplitude}.
    \textbf{a},	Spin expectation value, $\braket{\hat{I}_z}$, as a function of the RF pulse duration for 7 tones with equal amplitudes $V_{\rm RF}$. The solid lines are exponentially-decaying sinusoidal fits to the measured data points. 
    \textbf{b},	Rabi frequency $f_{\rm{Rabi}}^{\rm{CR}}$ extracted from the fits in \textbf{a} versus the RF driving amplitude $V_{\rm RF}$. 
    \textbf{c},	Numerically simulated error of covariant SU(2) rotations as a ratio of Rabi frequency and quadrupole splitting. Blue data points are obtained via a time evolution of Eq.~\eqref{eq:SuppCrossCouplingHamiltonian}. The red line represents a perturbative approximation. 
    The vertical dashed line represents the experimental setting used in the main text, where $V_{\rm RF}=22.86$ ~mV.}
    \label{fig:GRvsPower}
\end{figure}

The speed of covariant SU(2) rotations is tunable by changing the amplitude of the oscillating magnetic fields, $B_{1,k}$, applied to the nucleus. In the experiment, we tune $B_{1,k}$ by changing the amplitude of the output of the FPGA signal generator, which drives a current through the on-chip broadband antenna \cite{dehollain2012nanoscale} and generates the driving field.
In Fig.~\ref{fig:GRvsPower}a we show covariant Rabi oscillations for different output amplitudes of the source.
By fitting a sinusoidal decay to each Rabi oscillation, we extract the Rabi frequency $f_{\rm Rabi}^{\rm CR}$ for each driving amplitude (Fig.~\ref{fig:GRvsPower}b). We find a linear trend that is consistent with the expectation $f_{\rm Rabi}^{\rm CR} \propto B_{1,k}$.
 
Fig.~\ref{fig:GRvsPower}a, however, shows that the fidelity of the covariant Rabi oscillations deteriorates at higher driving amplitudes, i.e. when the Rabi frequency $f_{\rm Rabi}^{\rm CR}$ approaches the value of the quadrupole splitting $f_{\rm q}^+$. The consequent power broadening results in a cross-talk between different nuclear transitions. 
 
We determine the effect of power broadening by quantifying the decrease in contrast of the Rabi oscillations.
We model cross talk by introducing the term $\zeta = 1+e^{i\omega_\mathrm{q}t}+e^{i2\omega_\mathrm{q}t}+\dots +e^{i6\omega_\mathrm{q}t}$, with $\omega_\mathrm{q} = 2\pi f_{\rm q}^+$, which includes the interaction due to the resonant driving tone, and the $n = 6$ off-resonant terms that are separated by multiples of the quadrupole shift $f_{\rm q}^+$.
The driving Hamiltonian then yields
\begin{equation}
    \mathcal{\hat{H}}(t) = -\dfrac{ f_{\rm Rabi}^{\rm CR}}{4} \begin{pmatrix} 
    0 & \sqrt{7}\zeta& 0 & 0 & \dots & 0 \\
    
    \sqrt{7}\zeta^* & 0 & \sqrt{12}e^{-i\omega_\mathrm{q}t}\zeta & 0 & \dots & 0 \\
    \vdots & & \ddots & & & \vdots \\
    0 & 0 &  \dots  & & \sqrt{7}e^{-i6\omega_\mathrm{q}t}\zeta^* & 0 
    \end{pmatrix}\,.
    \label{eq:SuppCrossCouplingHamiltonian}
\end{equation}
We solve the time evolution of the system under Eq.~\eqref{eq:SuppCrossCouplingHamiltonian} starting from the state $\ket{-7/2}$ using QuTiP~\cite{johannson2013qutip}. We may solve either by directly integrating the time-dependent Schr\"odinger equation including these oscillatory terms, or perturbatively using average Hamiltonian theory, exploiting the fact that Eq.~\eqref{eq:SuppCrossCouplingHamiltonian} is periodic with period $T=1/f_q^+$.  For the latter strategy, Floquet's theorem assures that the solution to the Schr\"odinger equation for this periodic Hamiltonian is $\mathcal{U}_P(t)\exp(-i\mathcal{\hat{F}}t)$, where $\mathcal{\hat{U}}_P(t)$ is some perfectly periodic unitary (which resolves to identity every period), and $\mathcal{\hat{F}}$ is the Floquet hamiltonian, which we approximate via the Magnus expansion.  In the regime $f_{\rm{Rabi}}^{{\rm{CR}}} < f_q^+$ and to first order, $\mathcal{\hat{F}}\approx \bar{\mathcal{H}}^{(0)}+\bar{\mathcal{H}}^{(1)}$ where
\begin{align}
	\bar{\mathcal{H}}^{(0)} &= \frac{1}{T}\int_0^T dt_1 \mathcal{\hat{H}}(t_1) = \pi f_{\rm{Rabi}}^{\rm{CR}} I_x,\\
	\bar{\mathcal{H}}^{(1)} &= \frac{-i}{2T}\int_0^T dt_2 \int_0^{t_2} dt_1 [\mathcal{\hat{H}}(t_1),\mathcal{\hat{H}}(t_2)].
\end{align}
The first-order term $\bar{\mathcal{H}}^{(1)}$ is the correction to the simplest rotating-frame description of global Rabi driving given by $\bar{\mathcal{H}}^{(0)}$.  For the example of our $I=7/2$ system, this correction Hamiltonian evaluates to
\begin{equation}
\bar{\mathcal{H}}^{(1)}=
	\frac{(f^{\rm{CR}}_{\rm{Rabi}})^2}{16 f_{q}^+}
	\left(
	\begin{array}{cccccccc}
		-\frac{343}{20} & 0 & -\sqrt{\frac{7}{3}} & 0 & 0 & 0 & 0 & 0 \\
		0 & \frac{7}{4} & 0 & -\frac{6}{\sqrt{5}} & 0 & 0 & 0 & 0 \\
		-{\sqrt{\frac{7}{3}}} & 0 & \frac{133}{20} & 0 & -\sqrt{15} & 0 & 0 & 0 \\
		0 & -\frac{6}{\sqrt{5}} & 0 & \frac{35}{4} & 0 & -\sqrt{15} & 0 & 0 \\
		0 & 0 & -\sqrt{15} & 0 & \frac{35}{4} & 0 & -\frac{6}{\sqrt{5}} & 0 \\
		0 & 0 & 0 & -\sqrt{15} & 0 & \frac{133}{20} & 0 & -{\sqrt{\frac{7}{3}}} \\
		0 & 0 & 0 & 0 & -\frac{6}{\sqrt{5}} & 0 & \frac{7}{4} & 0 \\
		0 & 0 & 0 & 0 & 0 & -{\sqrt{\frac{7}{3}}} & 0 & -\frac{343}{20} \\
	\end{array}
	\right).
\end{equation}
\label{HGRF2}
 
Relative to a the simple SU(2) covariant rotation achieved by $\bar{\mathcal{H}}^{(0)}$, the correction from off-resonant terms causes an error scaling as $(f_{\rm{Rabi}}^{\rm{CR}}/f_q^+)^2$.  This means that a covariant rotation will lose fidelity as either the power is increased, or as the quadrupolar splitting $f_q^+$ decreases. Of course, in the opposite extreme of the perturbative expansion, $f_{\rm{Rabi}}^{\rm{CR}} \gg f_q^+$, we enter the ``hard-pulse" limit, in which a single-frequency broadband pulse can enact a SU(2) rotation to all transitions, and once again covariant operations are possible. The intermediate regime, $f_{\rm{Rabi}}^{\rm{CR}} \approx f_q^+$ is poorly described by perturbation theory. It corresponds to the regime where chaotic dynamics takes place, i.e. where the spin behaves as the quantum version of a chaotic periodically-driven top \cite{mourik2018exploring}.
 
Fig~\ref{fig:GRvsPower}c shows the normalised maximum peak height max($\braket{\hat{I}_z}$)$/I$ of a covariant Rabi oscillation including off-resonant terms, both using the average Hamiltonian above or direct integration of the time-dependent Schr\"odinger equation.  For very small Rabi frequencies, $f_{\rm{Rabi}}^{\rm{CR}} \ll f_q^+$, the effect is negligible and cross-coupling can be neglected.  As $f_{\rm{Rabi}}^{\rm{CR}}/f_q^+$ grows, Rabi contrast decreases, and displays chaotic behaviour around $f_{\rm{Rabi}}^{\rm{CR}}/f_q^+ \approx 1$.  After this regime, for $f_{\rm{Rabi}}^{\rm{CR}} \gg f_q^+$, covariant rotations are possible again as each frequency tone now drives all transitions.
Our experiment operates at $f_{\rm{Rabi}}^{\rm{CR}} \ll f_q^+ \approx 10^{-2}$, where the error due to cross-coupling is on the order of $\approx 10^{-4}$. If desiring global rotations at stronger $f_{\rm{Rabi}}^{\rm{CR}}$ and an elimination of this error, an appropriate strategy would be RF pulse shaping, which has been used effectively in the past to enact desired operations in quadrupole-split nuclear systems~\cite{mccoy_pulse_1992}.
 
\section{Creation of spin cat states with virtual-SNAP gate}
\label{SNAP}
\subsection{Arbitrary virtual-SNAP gates on the nuclear spin}
The selective number-dependent arbitrary phase (SNAP) gate \cite{Heeres2015} is designed to apply distinct phases $\xi_k$ to each eigenstate $\ket{k - 7/2}$ of $\hat{I}_z$, and it has the general form of 
\begin{equation}
    \hat{\mathcal{U}}_{\rm{SNAP}}
     \left( \Vec{\xi} \right) = \prod_{k=0}^{7} e^{i\xi_k \ket{k-7/2}\bra{k-7/2}}
\end{equation}
 
This gate originates from work in microwave oscillators, where an ancilla transmon qubit is used to introduce arbitrary phases to the oscillator's Fock states by performing transmon rotations that create a tunable geometric phase \cite{Heeres2015}. Our work differs from such setup in two important ways: (i) our system in intrinsically nonlinear (anharmonic) due to the nuclear quadrupole coupling, so we do not need to employ an ancilla qubit for state-selective operations, and (ii) we perform the SNAP gate virtually, by updating the phase of the GRF. This update is realized by a software instruction to shift by $\Delta \phi_k$ the phases of the 7 internal clocks in the Quantum Machines OPX+ signal generator. The driving terms in the laboratory frame Hamiltonian (Eq.~\eqref{HDS-eq:grf general Hamiltonian}) transform into: $B_{1,k}\cos(2\pi f_k t + \phi_k + \Delta \phi_k)$, which adds a phase shift to the off-diagonal terms in the generalised rotating frame Hamiltonian (for brevity, we set all initial $\phi_k$ to 0):
 
\begin{align}
\hspace{-1cm}
\hat{\mathcal{H}}_{\mathrm{update}} (\Vec{\Delta\phi}) = \frac{-\gamma_{\mathrm{n}}}{4}
  \left(
  \begin{array}{ccccccc}
  0 & \sqrt{7}B_{1,1}e^{i\Delta\phi_1} & 0 & \cdots & 0 & 0 \\
  \sqrt{7}B_{1,1}e^{-i\Delta\phi_1} & 0 & \sqrt{12}B_{1,2}e^{i\Delta\phi_2} & \cdots & 0 & 0 \\
  0 & \sqrt{12}B_{1,2}e^{-i\Delta\phi_2} & 0 & \cdots & 0 & 0 \\
  \vdots & \vdots & \vdots & \ddots & \vdots & \vdots \\
  0 & 0 & 0 & \cdots & 0 & \sqrt{7}B_{1,7}e^{i\Delta\phi_7} \\
  0 & 0 & 0 & \cdots & \sqrt{7}B_{1,7}e^{-i\Delta\phi_7} & 0 \\
  \end{array}
  \right)
\end{align}
 
\noindent where $\hat{\mathcal{H}}_{\rm{update}}$ denotes the Hamiltonian after the frame update and $\Vec{\Delta\phi} = (\Delta\phi_1, \Delta\phi_2, \cdots, \Delta\phi_7$) are the applied phase shifts to each of the 7 internal clocks. The question is now: what unitary has been applied to the state of the nucleus in the generalised rotating frame as a result of the frame rotation? We can see that in the interaction picture, the frame rotation transforms $\hat{\mathcal{H}}_{\rm{GRF}}$ as follows:
\begin{equation}
    \hat{\mathcal{H}}_{\rm{update}} \left( \Vec{\Delta\phi} \right) = \hat{\mathcal{U}}_{\rm{SNAP}} \hat{\mathcal{H}}_{\rm{GRF}} \hat{\mathcal{U}}_{\rm{SNAP}}^\dagger .
\end{equation}
 
\noindent where $\hat{\mathcal{U}}_{\rm{SNAP}} $ denotes the unitary SNAP gate: 
 
\begin{align}
\hat{\mathcal{U}}_{\mathrm{SNAP}}  =
\left(
\begin{array}{cccccccc}
1 & 0 & 0 & 0 & 0 & 0 & 0 & 0 \\
0 & e^{i\xi_1} & 0 & 0 & 0 & 0 & 0 & 0 \\
0 & 0 & e^{i\xi_2} & 0 & 0 & 0 & 0 & 0 \\
0 & 0 & 0 & e^{i\xi_3} & 0 & 0 & 0 & 0 \\
0 & 0 & 0 & 0 & e^{i\xi_4} & 0 & 0 & 0 \\
0 & 0 & 0 & 0 & 0 & e^{i\xi_5} & 0 & 0 \\
0 & 0 & 0 & 0 & 0 & 0 & e^{i\xi_6} & 0 \\
0 & 0 & 0 & 0 & 0 & 0 & 0 & e^{i\xi_7} \\
\end{array}
\right)
\end{align}
 
\noindent We can now express the phase gate angles $\xi_k$ into the frame rotation angles $\Delta \phi_k$ as: 
\begin{equation}
    \xi_k = - \sum_{i=1}^k{\Delta \phi_i}
    \label{phase_update}
\end{equation}
 
\noindent This relationship demonstrates that the phase acquired on the $\hat{I_z}$ eigenstate $\ket{k - 7/2}$ is the cumulative sum of the frame update angles from $\Delta\phi_1$ to $\Delta\phi_k$. This approach provides a straightforward and nearly instantaneous (within a single clock cycle of the OPX+, 4~ns) method to implement a SNAP gate. Although it's easiest to recognize the effect of the frame update by writing it in terms of the driving Hamiltonian, this SNAP gate is \emph{virtual} because it does not require any signal to be physically applied to the nucleus. 
 
\subsection{Implementation of one-axis twisting by virtual-SNAP}
The second-order nonlinearity, $\hat{I}_z^2$, in the Hamiltonian (Eq.~1 of the main text) introduces one-axis twisting of the spin state, a widely studied phenomenon in the context of spin squeezing \cite{Kitagawa_1993}. In the laboratory frame, the one-axis twisting dynamics can induce the revival and collapse between the Schr\"{o}dinger cat state and the spin coherent state lying on the equatorial plane of the Bloch sphere \cite{Gupta2024}. An appealing way to conceptualise the ability of the virtual-SNAP gate to suddenly transform a spin coherent state into a cat state is by realising that such transformation happens anyway, periodically, in the laboratory frame. The initial $\ket{\rm scs_{7/2}}_x$ state only appears static because we have `locked' onto it through the definition of the GRF. By switching to a different set of GRF phases through the virtual-SNAP, we suddenly 'lock' to a different point along the collapse-revival dynamics that takes place in the lab frame, and we choose that point to be the one where a cat state is formed.
 
In particular,
\begin{align}
R_{\pi/2}(0) e^{-i I_z^2\pi/2} R_{\pi/2}(-\pi/2) \equiv 
\frac{1}{\sqrt{2}}
\left(
\begin{array}{cccccccc}
1 &0 &0& 0&0&0&0&i\\
0 &-i&0& 0&0&0&1&0\\
0 & 0&1& 0&0& i&0&0\\
0 & 0&0&-i &1& 0&0&0\\
0 & 0&0& i &1& 0&0&0\\
0 & 0&1& 0&0&-i&0&0\\
0 & i&0& 0&0&0&1&0\\
1 &0 &0& 0&0&0&0&-i\\
\end{array}
\right),
\end{align}
with equivalence up to virtual-SNAPs, which clearly enables the construction of cat-states in any subspace using covariant rotations and $\hat{I}_z^2$ evolution only.
Given the fact that $\hat{I}_z^2$ exclusively modifies phases while leaving state population unaffected, we are able to utilise the virtual-SNAP gate to implement the one-axis twisting by adjusting phases in the generalised rotating frame. However, it should be noted that by using the virtual-SNAP gate, one can only generate certain equatorial cat states -- those with identical populations to spin coherent state along the equator. 
 
In practice, we perform a covariant SU(2) rotation to the initial eigenstate $\ket{-7/2}$ to first prepare the coherent state $\ket{\rm scs_{7/2}}_{-x}$, and then apply $\hat{\mathcal{U}}_{\rm{SNAP-cat}}$ to create an $x$-oriented Schr\"{o}dinger cat state $\ket{\mathrm{cat}_{7/2}, \xi_7 = \pi/2}_x$, with the form:  
\begin{subequations}
\begin{minipage}{0.5\linewidth}
\begin{equation}
\ket{\rm scs_{7/2}}_{-x} = 
  \left(
  \begin{array}{r}
    -0.088 \\
    0.234 \\
    -0.405 \\
    0.523 \\
    -0.523 \\
    0.405 \\
    -0.234 \\
    0.088 \\
  \end{array}
  \right)
\end{equation}
\end{minipage}
\hspace*{\fill}
\begin{minipage}{0.5\linewidth}
\begin{equation}
  \ket{\mathrm{cat}_{7/2}, \xi_7 = \pi/2}_x = 
  \left(
  \begin{array}{cc}
    -0.088 \times e^{-i\frac{3\pi}{4}} \\
    0.234 \times e^{-i\frac{\pi}{4}} \\
    -0.405 \times e^{-i\frac{3\pi}{4}} \\
    0.523 \times e^{-i\frac{\pi}{4}} \\
    -0.523 \times e^{-i\frac{3\pi}{4}} \\
    0.405 \times e^{-i\frac{\pi}{4}} \\
    -0.234 \times e^{-i\frac{3\pi}{4}} \\
    0.088 \times e^{-i\frac{\pi}{4}} \\
  \end{array}
  \right)
\end{equation}
\end{minipage}
\label{cat_x_create}
\end{subequations}
 
It is clear that $\hat{\mathcal{U}}_{\rm{SNAP-cat}}$ with the following form is able to induce the transition from $\ket{\rm scs_{7/2}}_{-x}$ to $\ket{\mathrm{cat}_{7/2}, \xi_7 = \pi/2}_x$. 
\begin{equation}
\hat{\mathcal{U}}_{\rm{SNAP-cat}} = e^{-i\frac{3\pi}{4}}
\left( \begin{array}{cccccccc}
1 & 0 & 0 & 0 & 0 & 0 & 0 & 0 \\
0 & e^{i\frac{\pi}{2}} & 0 & 0 & 0 & 0 & 0 & 0 \\
0 & 0 & 1 & 0 & 0 & 0 & 0 & 0 \\
0 & 0 & 0 & e^{i\frac{\pi}{2}} & 0 & 0 & 0 & 0 \\
0 & 0 & 0 & 0 & 1 & 0 & 0 & 0 \\
0 & 0 & 0 & 0 & 0 & e^{i\frac{\pi}{2}} & 0 & 0 \\
0 & 0 & 0 & 0 & 0 & 0 & 1 & 0 \\
0 & 0 & 0 & 0 & 0 & 0 & 0 & e^{i\frac{\pi}{2}} \\
\end{array} \right)
\end{equation}
Based upon Eq.~\eqref{phase_update}, this  $\hat{\mathcal{U}}_{\rm{SNAP-cat}}$ gate can be in practice achieved in the GRF with a software instruction to update:
\begin{equation}
    \Vec{\Delta\phi} = (-\pi/2, +\pi/2, -\pi/2, +\pi/2,-\pi/2, +\pi/2,-\pi/2)
\end{equation}
The method of applying alternating phases of $-\pi/2$ and $\pi/2$ is applicable more generally to create zero-parity $\ket{\mathrm{cat}_{I}}_x$ for all half-odd integer spins. We thus employ this method to create subspace cats ($\ket{\rm cat_{5/2}}_x$, $\ket{\rm cat_{3/2}}_x$) as illustrated in Fig.~4e of the main text.
 
\section{Subspace rotations and different size cat states}
\label{SI_sec:subRot}
 
In the nuclear spin-7/2 system, it is possible to restrict the full $8$-dimensional Hilbert space to specific subspaces, thereby simulating spins of varying sizes. For example, by constraining to the subspace spanned by $\{\ket{3/2}, \ket{1/2}, \ket{-1/2}, \ket{-3/2}\}$, we effectively emulate a spin-3/2 system. 
 
The covariant SU(2) rotations in the subspace require adjusting the driving amplitude, $B_{1,k}$, of each tone. One must ensure that the driving strength, determined by $B_{1,k}$ and the element of the spin-7/2 operator $\hat{I}_x^{7/2}$, matches the spin operator in the subspace. For instance, one can implement a spin-3/2 covariant rotation by matching $\sqrt{15}B_{1,3}$, $\sqrt{16}B_{1,4}$, and $\sqrt{15}B_{1,5}$, to the corresponding off-diagonal elements in the spin-3/2 $\hat{I}_x^{3/2}$ operator, while all other $B_{1,k}$ are set to zero. The corresponding spin operator $\hat{I}_x^{3/2}$ for the spin-3/2 system has the form:
\begin{equation}
\hat{I}_x^{3/2}  = \frac{1}{2}
\left(
\begin{array}{cccc}
0 & \sqrt{3} & 0 & 0 \\
\sqrt{3} & 0 & 2 & 0 \\
0 & 2 & 0 & \sqrt{3} \\
0 & 0 & \sqrt{3} & 0
\end{array}
\right).
\end{equation}
 
\begin{figure}[ht!]
    \centering
    \includegraphics{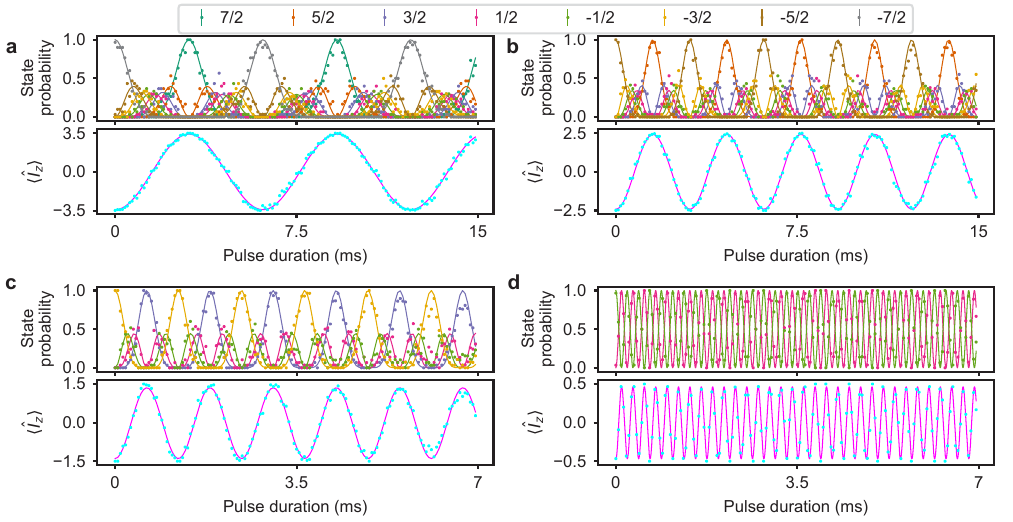}
    \caption{\textbf{Subspace covariant SU(2) rotations}. Subspaces with sizes \textbf{a} $I = 7/2$, \textbf{b} $I = 5/2$, \textbf{c} $I = 3/2$, \textbf{d} $I = 1/2$. Each panel shows the evolution of the state probabilities (top) and of the $\langle \hat{I}_z \rangle$ expectation values (bottom).
    Cyan points show $\langle \hat{I}_z \rangle$ expectation values calculated from the state probabilities. Solid magenta lines show a sinusoidal fit to extract the Rabi frequency.}
    \label{fig:SubspaceRotations}
\end{figure}

The experimental demonstration of subspace covariant SU(2) rotations is shown in Fig.~\ref{fig:SubspaceRotations}.
We initialise the spin in the $\hat{I}_z$ eigenstate with the highest energy, e.g. $\ket{-7/2}$ for a spin-7/2 subspace, $\ket{-3/2}$ for the spin-3/2 subspace.
A multi-frequency NMR pulse is applied with the power of each tone adjusted to implement a subspace rotation.
For each subspace, the $\langle \hat{I}_z \rangle$ expectation value shows a sinusoidal oscillation with an amplitude that corresponds to the size of the subspace.
From these oscillations, we extract the Rabi frequency of the subspace covariant SU(2) rotation via a sinusoidal fit. As the total power is held constant, the Rabi frequency increases as the subspace gets smaller.
We then use the Rabi frequency to numerically simulate an ideal subspace rotation around the $-y$-axis and find excellent agreement with our data.
 
\begin{figure}[ht]
  \centering
  \includegraphics{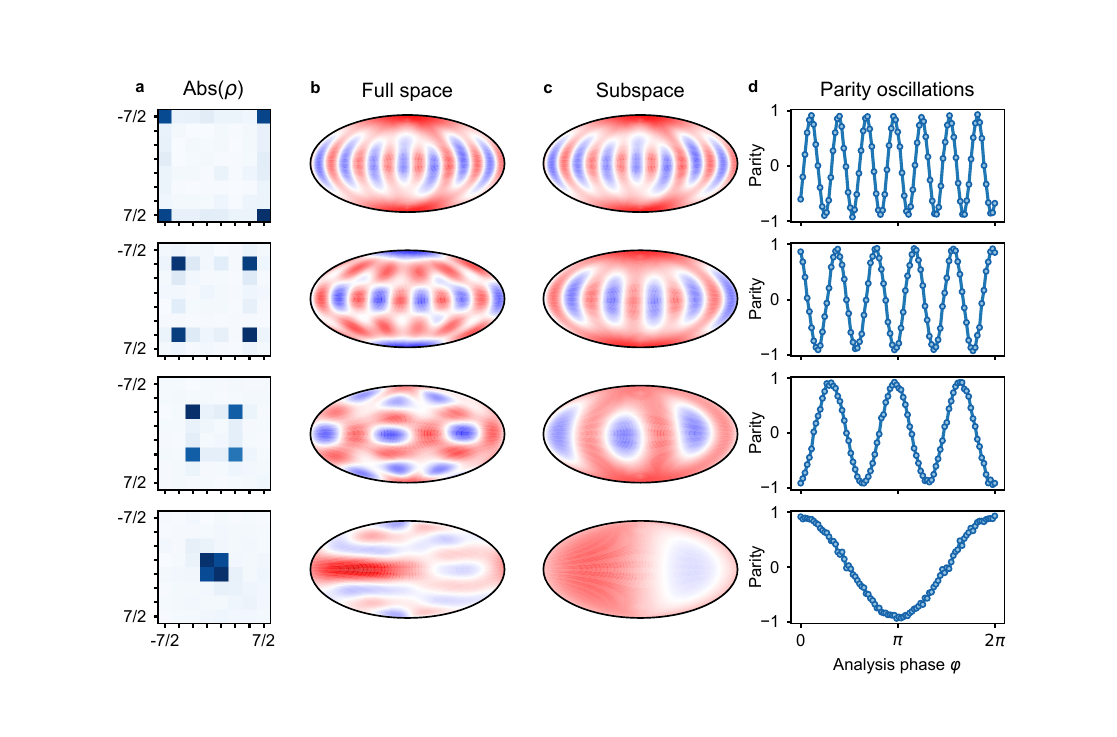}
  \caption{
  \textbf{Full and subspace Schr\"{o}dinger cat states in the $I=7/2$ nuclear spin system}.
  Visualisation of density matrix reconstruction and parity oscillations for cat states across dimensions $d=8$ (top row), 6 (second row), 4 (third row), and 2 (bottom row). The panels depict: \textbf{a}, absolute value of the density matrix, \textbf{b}, Wigner function in the 8-dimensional Hilbert space, \textbf{c}, Wigner function of the truncated density matrix highlighting cat-like properties, and \textbf{d}, parity oscillations, showing $d-1$ periods.}
  \label{fig:different_size_cats}
\end{figure}

Fig.~\ref{fig:different_size_cats} illustrates the Wigner functions of spin Schr\"{o}dinger cat states within subspaces of dimensions d=8, 6, 4, and 2 (the $d=2$ case is not a cat state by any sensible definition, but we show it here to illustrate the universality of the operations), with corresponding density matrices and parity oscillations. The different-sized cat states are created using the Givens rotations protocol outlined in Fig.~3a-b of the main text. Subsequently, we perform density matrix tomography using the method described in Section~\ref{Density matrix reconstruction}. Fig.~\ref{fig:different_size_cats}a shows the absolute value of the density matrix of cat states in four different subspaces, sorted in descending order by size. Fig.~\ref{fig:different_size_cats}b shows the Wigner function of the reconstructed density matrix in the full 8-dimensional Hilbert space. To highlight the cat-like features of the produced states, we truncate and normalize the reconstructed density matrix and plot the Wigner function of the resulting reduced density matrix in Fig.~\ref{fig:different_size_cats}c. It becomes clear that the number of peaks and valleys in the Wigner function scales with the dimension of the cat state. For instance, in the spin-7/2 cat, there are 7 valleys; in the spin-5/2 cat, there are 5 valleys, and so on. This scaling is also reflected in Fig.~\ref{fig:different_size_cats}d, where we display the parity oscillations of the different size cat state using the same method as in Fig.~3 of the main text. The number of periods of a cat state spanning $d$ dimensions is $d-1$. The appearance of Wigner function negativity even in the trivial spin-1/2 case may seem surprising at first sight, but it reflects the fact that even Gaussian-like spin coherent states (which all pure spin-1/2 states are forced to be \cite{Kitagawa_1993}) have Wigner negativity \cite{davis2023stellar}, unless $I\rightarrow \infty$.
 
\newpage
\section{Parity measurement}
\label{SI_sec:parity}
\def\ds{\rule{0pt}{1.5ex}}
In order to determine the coherence and fidelity of the $\ket{\rm cat_{7/2}}_z$ state, we use a reduced tomography method based on measuring parity oscillations \cite{Turchette_1998}.
The state created in the experiment is described by the density matrix, $\rho$, with elements $\rho_{i,j}$, where $i,j$ denote the $m_I$ spin projections onto the $z$-axis.
The fidelity, $\mathcal{F}$, between $\rho$ and the cat state $\ket{\rm cat_{7/2}}_z = (\ket{7/2}+e^{i\xi_7}\ket{-7/2})/\sqrt{2}$ is given by
\begin{equation}
    \mathcal{F} ={}_{\ds z}
 \braket{\rm cat_{7/2}|\rho|\rm cat_{7/2}}_{z}
    = \dfrac{1}{2}(\rho_{-7/2,-7/2}+\rho_{7/2,7/2}) + \dfrac{1}{2}(e^{i\xi_7}\rho_{-7/2,7/2}+\rm{c.c.}),
    \label{eq:SuppCatStateFidelity}
\end{equation}
where the first term corresponds to the sum of the state probabilities of $\ket{-7/2}$ and $\ket{7/2}$, and the second term to the sum of the coherences. The state probabilities are determined with a projective measurement in the $\hat{I}_z$ basis, while the coherences are determined by rotating the $\ket{\rm cat_{7/2}}_z$ state to the equator of the spin sphere by applying a covariant SU(2) rotation, $R_{\pi/2}(\varphi)$.
 
We then evaluate the parity operator $\hat{\Pi} = \sum_{m_I} (-1)^{I+m_I}\ket{m_I}\bra{m_I}$ as a function of $\varphi$ and find
\begin{equation}
    \langle \hat{\Pi}(\varphi) \rangle =  \dfrac{1}{2}\sum_{m_I}(-1)^{I+m_I}(e^{i2m_I\varphi}\rho_{m_I,-m_I}-e^{-i2m_I\varphi}\rho_{-m_I,m_I}),
\end{equation}
which describes a sum of sinusoidal oscillations with periods $2\pi/7$, $2\pi/5$, $2\pi/3$, $2\pi$ and amplitudes depending on the absolute value of the coherences.
For the $\ket{\rm cat_{7/2}}_z$ state prepared in the experiment, we find $\rho_{\pm7/2, \pm7/2}\gg \rho_{ i, i}$ with $i = \pm5/2, \pm3/2, \pm1/2$ and can therefore neglect the oscillations with periods $>2\pi/7$ as we expect the coherences between the unpopulated states to be zero.
The remaining oscillation that we observe in the experiment (see main text Fig.~2c,g) is then dominated by the term $(e^{i7\varphi}\rho_{-7/2, 7/2} + \rm{c.c.})$, which corresponds to the second term of Eq.~\eqref{eq:SuppCatStateFidelity}.
Hence, the state fidelity can then be computed by averaging the state populations and the contrast of the parity oscillation. In addition, the phase offset of the parity oscillation gives information about the phase $\xi_7$ of the state.
 
\section{Density matrix reconstruction}
 
\label{Density matrix reconstruction}
It is \textit{possible} to estimate the Wigner function $W(\theta,\varphi)$ of a large spin by measuring it directly at a dense grid of points $(\theta_n,\varphi_n)$, expressing $W(\theta,\varphi)$ in terms of the outcome probabilities of a spin measurement in the $(\theta_n,\varphi_n)$ direction \cite{chen2019quantum}. However, a much more efficient protocol is to estimate the Wigner function by using Eq.~\eqref{spin Wigner 1} to relate $W(\theta,\varphi)$ to a $(2I+1)\times (2I+1)$ density matrix $\rho$.  Unlike $W(\theta,\varphi)$, which is defined at uncountably many points, $\rho$ is described by just $4I(I+1)$ numbers, and can be estimated using quantum state tomography \cite{hradil1997quantum}.  We do so, and then compute $W_\rho(\theta,\varphi)$ at all points using the tomographic estimate $\rho_{\rm MLE}$.
 
\subsection{Tomographic measurement protocol}

At least $2I+2$ different quantum measurement bases are required for tomography.  Each basis measurement has $2I+1$ distinct outcomes whose probabilities yield $2I$ independent real numbers, so $2I(2I+2) = 4I(I+1)$ density matrix elements can be deduced.  The measurements we can make easily are measurements of $\mathbf{\hat{I}}\cdot\mathbf{n}$, where $\mathbf{n}\sim(\theta,\varphi)$ is a unit vector .  Because $\mathbf{\hat{I}}\cdot\mathbf{n}$ measurements are not entirely uncorrelated with each other, measuring along $2I+2$ axes is not sufficient.  Instead, $4I+1$ are both necessary \cite{hofmann2004quantum} and sufficient \cite{newton1968measurability}.  For our spin-$7/2$ system, $4I+1=15$.  But achieving optimal efficiency (accuracy vs number of samples) does not appear to be possible with a minimal set of $4I+1$ measurements \cite{perlin2021spin}.
 
The tomographic efficiency of a particular set of measurements (a.k.a.~\textit{experiment design}) is determined by the condition number of the \textit{frame superoperator} whose action on a state $\rho$ is
\begin{equation}
    F[\rho] = \frac{1}{N}\sum_{j=1}^N{ E_j \Tr[ E_j] },
\end{equation}
where $E_j$ denotes an \textit{effect} of a measurement (i.e., the projector $\proj{m_{\theta_n,\varphi_n}}$ onto an eigenstate of $\mathbf{\hat{I}}\cdot\mathbf{n}_{\theta_n,\varphi_n}$), and ranges over all $N$ effects in the experiment design (every effect of every measurement performed).  If $F$ is rank-deficient, then the experiment design is not \textit{informationally complete}, and precise tomography is impossible.  A good experimental design is one for which the spectrum of $F$ is as flat as possible, and has no small eigenvalues.  A good metric for quantifying tomographic efficiency is 
\begin{equation}
    f_{\rm te}(F) = \sqrt{\Tr(F^{-1})},
\end{equation}
with \emph{smaller} $f_{\rm te}$ being better.  The best possible tomographic efficiency for $I=7/2$, achieved by a 2-design \cite{renes2004symmetric}, is $f_{\rm te} = \sqrt{4545} \approx 67.41$.
 
Among experiment designs comprised entirely of $\mathbf{\hat{I}}\cdot\mathbf{n}$ measurements, the best performance is achieved by measuring \textit{all} axes according to a spherically uniform distribution ($f_{\rm te} = \sqrt{5440} \approx 73.76$).  This is obviously impossible in an experiment, since there are infinitely many axes.  We used numerical analysis to find a small experiment design comprising just $45 = 3 \times 15$ measurement axes whose efficiency is very close to optimal ($f_{\rm te} \approx 76.3$).  This experiment design comprises a grid over $(\theta,\varphi)$, with 15 equally spaced values of $\varphi_n = n\frac{2\pi}{15}$ ($n=0\ldots 14$), and 3 specific values of $\theta_n = \{\frac{\pi}{4}, \frac{\pi}{3}, \frac{9\pi}{20}\}$, where $\theta=0$ indicates the sphere's axis and $\theta=\pi/2$ its equator.  
 
We used this experiment design for all state reconstructions.  In most cases, we performed 15 shots of each measurement, for a total of $15 \times 45 = 675$ shots.  In one case we performed 50 shots of each measurement, and in two cases we performed 80 shots.
 
\subsection{Maximum likelihood state estimation}
 
For each state that we reconstructed, we recorded all the data (outcomes of each shot), and then used unmodified maximum likelihood estimation (MLE) \cite{hradil1997quantum} to obtain an estimate $\rho_{\rm MLE}$ of the density matrix. 
The data form a list $\{(E_j, n_j)\}$ of effects ($E_j = \proj{m_{\theta_j,\varphi_j}}$) and their observed frequencies ($n_j$). 
The MLE is obtained by constructing the \textit{likelihood function},
\begin{equation}
    \mathcal{L}(\rho) \equiv \mathrm{Pr}(\mathrm{data}|\rho) = \prod_j{\Tr(E_j\rho)^{n_j}},
\end{equation}
and then finding its maximum over all $\rho$ satisfying $\rho\geq0$ and $\Tr\rho=1$.
In practice, we minimize the negative \textit{loglikelihood} $-\log(\mathcal{L})$ because it is convex.
 
We implemented this method with CVXPY 1.4 as a modeling platform and MOSEK 10 as a numerical solver \cite{cvxpy2018,Dahl2021}.
Specifically, letting \texttt{Es} denote a matrix whose rows are obtained by flattening each $E_j$ in row-major order, and letting \texttt{ns} denote a corresponding vector of observed frequencies, the MLE is obtained as follows:
\begin{verbatim}
    import cvxpy as cp
    rho = cp.Variable(shape=(8, 8), hermitian=True)
    probs = cp.real(Es @ rho.flatten(order='C'))
    loglikelihood = cp.sum(cp.multiply(ns, cp.log(probs)))
    objective = cp.Maximize(loglikelihood)
    problem = cp.Problem(objective, [rho >> 0, cp.trace(rho) == 1])
    problem.solve(solver='MOSEK')
    rho_MLE = rho.value
\end{verbatim}
We ran this code on many different datasets (both real and simulated) and found that it took between 0.1 and 0.2 seconds to execute on a MacBook Pro with an M2 Max processor.
Notably, this is less than the time required just to compute the eigenstates of $\mathbf{\hat{I}}\cdot\mathbf{n}_{\theta_n,\varphi_n}$ by standard numerical methods.
Therefore for improved efficiency we computed these eigenstates by leveraging a symbolic eigendecomposition of $\mathbf{\hat{I}}\cdot\mathbf{n}_{\theta_n,\varphi_n}$ obtained with Maple.
 
\begin{figure}[ht!]
    \centering
    \includegraphics[width=\columnwidth]{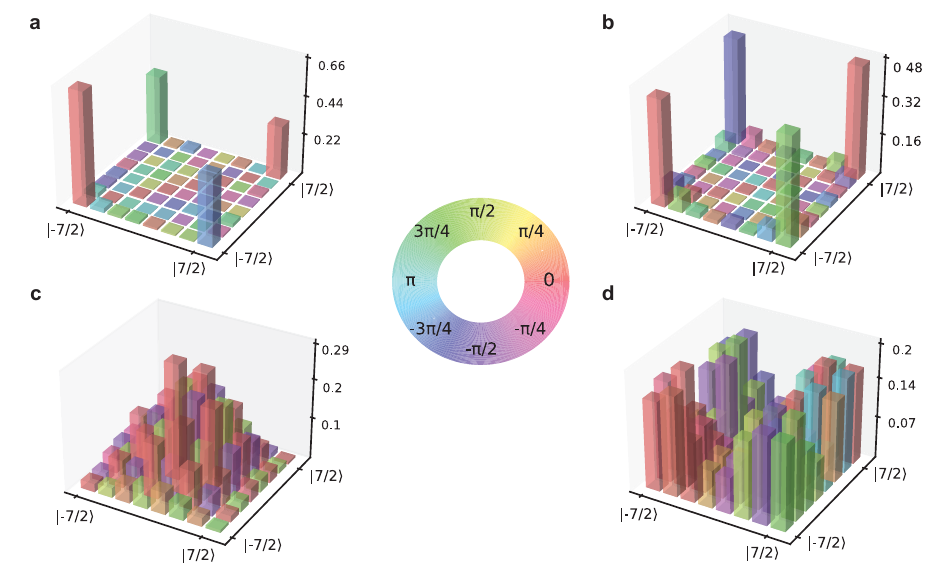}
    \caption{\textbf{Reconstructed density matrices of the spin cat states}.
    \textbf{a}, $\ket{\mathrm{cat}_{7/2}, \xi_7 = -\pi}_z = (\ket{-7/2}-\ket{7/2})/\sqrt{2}$, prepared by Givens rotations.
    \textbf{b}, $\ket{\mathrm{cat}_{7/2}, \xi_7 = \pi/2}_z = (\ket{-7/2}+i\ket{7/2})/\sqrt{2}$, prepared by the sequence of $\pi/2$ covariant SU(2) rotation + virtual-SNAP + $\pi/2$ covariant SU(2) rotation.
    \textbf{c}, $\ket{\mathrm{cat}_{7/2}, \xi_7 = \pi/2}_x$, prepared by the sequence of $\pi/2$ covariant SU(2) rotation + virtual-SNAP.
    \textbf{d} Cat state obtained after a $\pi/4$ covariant SU(2) rotation of $\ket{\mathrm{cat}_{7/2}, \xi_7 = \pi/2}_x$, i.e. the second-last example depicted in the Hammer projection of Fig.~3h of the main text.
    }
    \label{fig:DensityMatrix}
\end{figure}
 
\subsection{Confirming the validity of the state tomography}
 
State tomography has known weaknesses, because it relies on knowing exactly what measurements were performed.  If the measurements actually performed in the experiment were noisy, flawed, or otherwise different from the mathematical models $\{E_n\}$ used in the MLE reconstruction, then the estimate $\rho_{\rm MLE}$ will be systematically wrong.  We sought to rule out as many potential causes of systematic mis-estimation as possible.
 
First, we consider the possibility that we might have been performing \textit{noisy} measurements, whose effects $\{E_n\}$ are not rank-1 projectors.  Although this cannot be ruled out, we can say that its impact is small.  We modeled our measurement effects as perfect rank-1 projectors.  Additional noise in all the measurements would look the same as noise in the state, and would cause $\rho_{\rm MLE}$ to look more mixed (less pure).  We do \textit{not} observe this -- our estimated states always have low rank, and are nearly pure.  This certifies that our measurement effects are indeed close to rank-1.
Second, we consider the possibility that the measurement axes might be systematically rotated by some SU(2) unitary.  This also cannot be ruled out.  However, it is also entirely inconsequential.  Our main results are about the creation of ``Schr\"{o}dinger's Cat" states that are highly delocalised over the sphere, and those properties are invariant under SU(2) rotations.  So systematic rotations of the measurement axes have no effect on our results.
 
\begin{figure}[t!]
    \centering
    \includegraphics[scale=0.75,angle=90]{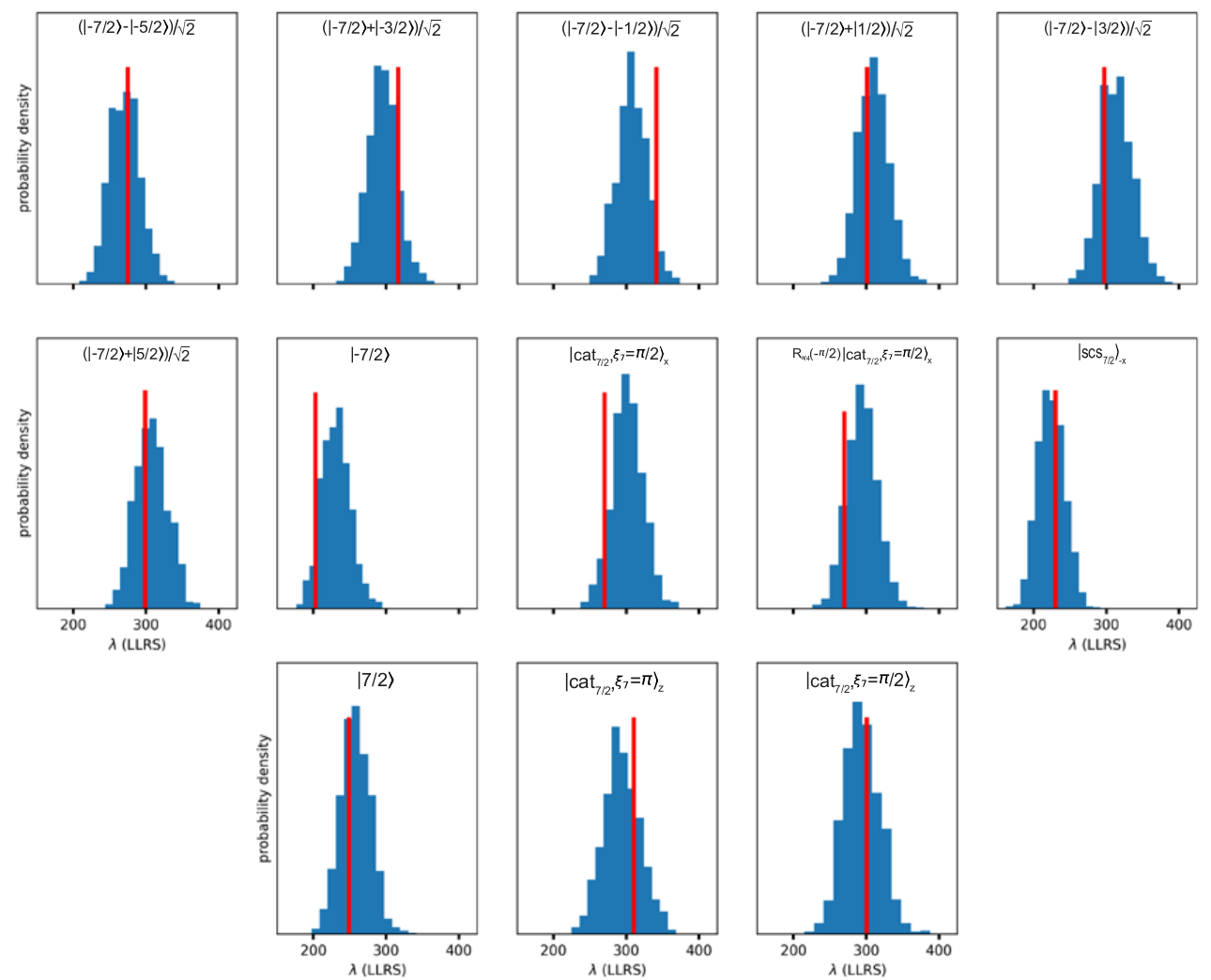}
    \caption{\textbf{Validation of tomographic state reconstructions}.  To check for possible errors or corruption in the measurement axes, a.k.a.~\emph{model violation}, we computed the loglikelihood ratio statistic ($\lambda$, see text for details) between the MLE state $\rho_{\rm MLE}$ and a saturated model.  For each of the 13 states reconstructed, we used a parametric bootstrap to generate 1000 samples from the null distribution of $\lambda$ (blue histogram) and compared the experimental $\lambda$ (red) to it.  No inconsistency was observed. The total runtime for computing the 24,000 MLEs needed by this procedure was under one hour.}
    \label{fig:TomographyValidation}
\end{figure}
 
\begin{table}[ht]
\centering
\begin{tabular}{l>{\centering\arraybackslash}p{0.2\linewidth}>{\centering\arraybackslash}p{0.2\linewidth}}
\toprule
Target State & Method of Creation & Fidelity \\
\midrule
 $\ket{-7/2}$ & Initial state & 0.990(1)\\
 $\ket{\mathrm{scs}_{7/2}}_{-x}$& CR & 0.934(3)\\
 $\ket{7/2}$ & CR & 0.890(3)\\
$(\ket{-7/2}-\ket{-5/2})/\sqrt{2}$ & Given rotations & 0.909(2)\\
$(\ket{-7/2}+\ket{-3/2})/\sqrt{2}$ & Given rotations & 0.819(4)\\
$(\ket{-7/2}-\ket{-1/2})/\sqrt{2}$ & Given rotations & 0.847(3)\\
$(\ket{-7/2}+\ket{1/2})/\sqrt{2}$ & Given rotations & 0.879(3)\\
$(\ket{-7/2}-\ket{3/2})/\sqrt{2}$ & Given rotations & 0.780(4)\\
$(\ket{-7/2}+\ket{5/2})/\sqrt{2}$ & Given rotations & 0.866(3)\\
$\ket{\mathrm{cat}_{7/2}, \xi_7 = \pi/2}_x$ & CR + virtual-SNAP & 0.883(4)\\
$R_{\pi/4}(-\pi/2)\ket{\mathrm{cat}_{7/2}, \xi_7 = \pi/2}_x$ & CR + virtual-SNAP & 0.917(3)\\
$(\ket{-7/2}-\ket{7/2})/\sqrt{2}$ & Givens rotations & 0.794(2)\\
$(\ket{-7/2}+i\ket{7/2})/\sqrt{2}$ & CR + virtual-SNAP & 0.874(2)\\
\bottomrule
\end{tabular}
\caption{State fidelity, $\mathcal{F} = \braket{\psi|\rho_{\rm MLE}|\psi}$, where $\psi$ is the target state, and $\rho_{\rm MLE}$ is the density matrix obtained from maximum likelihood quantum state tomography.
}
\label{tab:fidelity}
\end{table}
 
Finally, we consider the possibility that \textit{some} measurement axes might be erroneously rotated relative to the rest.  This is the most interesting possibility.  If the relative orientation of the measurement axes were to be corrupted in this way, it would cause inconsistency within the dataset \cite{shulman2012demonstration,van2013quantum}.  We can test for this using a loglikelihood ratio test \cite{blume2010entanglement} that compares the loglikelihood (a measure of goodness-of-fit) of the MLE density matrix $\rho_{\rm MLE}$ to a \textit{saturated model} that is allowed to fit every measurement independently \cite{nielsen2021gate}.  In the absence of any measurement-axis inconsistencies, the \textit{null distribution} of the loglikelihood ratio statistic,
\begin{equation}
    \lambda = -2\log\left(\frac{\mathcal{L}(\rho_{\rm MLE})}{\mathcal{L}_{\mathrm{saturated}}}\right)
\end{equation}
can be computed.  Under ideal conditions, $\lambda$ would behave as a $\chi^2_k$ random variable with $k = 315-63 = 252$ degrees of freedom, because the maximal model has 315 degrees of freedom and the density matrix has 63.  However, positivity constraints on $\rho$ complicate this picture \cite{scholten2018behavior}.  Therefore, we used a parametric bootstrap (1000 samples) to evaluate the null distribution of $\lambda$ for each of the 24 states we reconstructed.  Then, we computed the experimental value of $\lambda$ for each state, and compared them.  The results are shown in Fig.~\ref{fig:TomographyValidation}.  For each state, the observed $\lambda$ is consistent with the null distribution, meaning that there is \textit{no} evidence for any significant errors in the measurements.
 
In summary, we used the most powerful available statistical tools to search for evidence of inconsistencies in our tomography, and found none. 
 
\newpage
\section{$T_2^{\ast}$ for adjacent nuclear spin levels}
\label{SI_sec:T2Star}
We have repeated on this device an experiment reported earlier in Ref.~\cite{fernandez2024navigating}, where we measured the nuclear dephasing times $T_2^{\ast}$ by Ramsey experiments involving adjacent nuclear levels, i.e. by creating superpositions of the form $(\ket{m_I}+\ket{m_I -1})/\sqrt{2}$. We found again the expected trend where $T_2^{\ast}(m_I \leftrightarrow m_I - 1)$ is maximum for the $\ket{1/2} \leftrightarrow\ket{-1/2}$ transition, since it is to first-order insensitive to fluctuations in the nuclear quadrupole splitting, which can be caused by electrical noise in the device. The value of $T_2^{\ast}(1/2 \leftrightarrow -1/2) = 167(60)$~ms is particularly high in this device, which could be the result of a fortunate absence of $^{29}$Si nuclear spins in the immediate vicinity of the $^{123}$Sb nucleus under study (see Section~\ref{sec:other_devices} for counterexamples).
 
\begin{figure}[ht!]
    \centering
    \includegraphics[width=12cm] {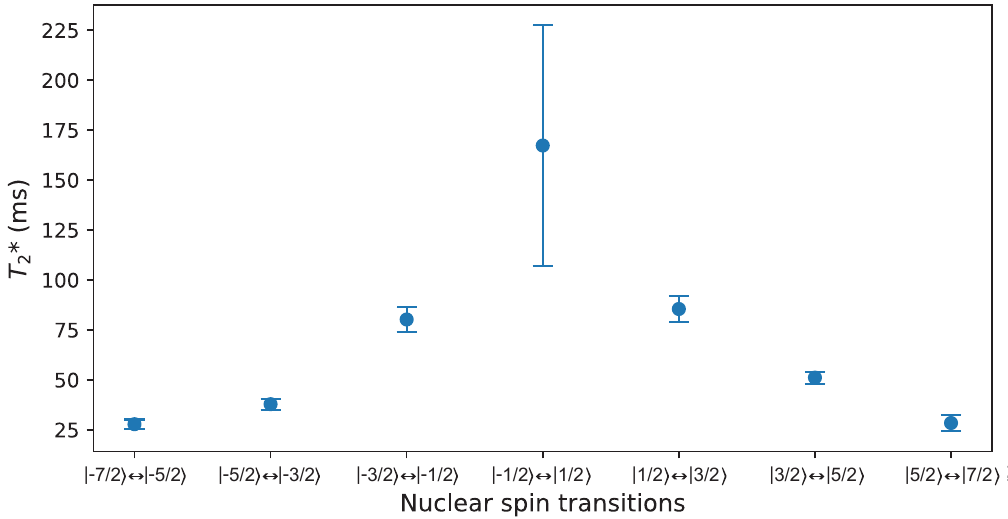}
    \caption{
    \textbf{$\mathbf{T_2^{\ast}}$ for adjacent spin levels}. The value of
    $T_2^{\ast}$ is extracted from fitting the Ramsey free induction decay. The total experiment time of each Ramsey measurement is kept identical to ensure that the system experiences consistent temporal noise features.
    }
    \label{fig:ramsey}
\end{figure}
 
\section{Nuclear spin readout and initialisation}
\label{sec:nuclear_readout}
 
The initialisation and readout of the nuclear spin follows the methods described in \cite{asaad2020coherent,fernandez2024navigating}. We summarise them again here for the readers' convenience, and to highlight the key features that affect the way we perform the experiments described in the main text.
 
Our experiments on creation and manipulation of Schr\"{o}dinger cat states focus on using the ionised $^{123}$Sb nuclear spin, i.e. the donor in the charge-positive $D^+$ state. However, reading out the nuclear spin populations requires introducing a hyperfine-coupled electron spin, which acts as a readout ancilla, as first demonstrated in the simpler $^{31}$P system \cite{pla2013high}. The electron is reintroduced on the donor by adjusting the voltages on the donor gates. 
 
The Hamiltonian of the charge-neutral ($D^0$), coupled electron-nuclear system takes the form:
\begin{equation}
    \hat{\mathcal{H}}_{D^0} = \gamma_{\rm{e}}B_{0}\hat{S_{z}} - \gamma_\mathrm{n} B_{0}\hat{I_{z}} + A \cdot \hat{\textbf {S}}\cdot \hat{\textbf{I}} + \hat{\mathcal{H}}_{Q^0},
    \label{neutral Hamiltonian}
\end{equation}
where $\gamma_{\rm{e}} = 27.97$~GHz/T is the electron gyromagnetic ratio , $\gamma_\mathrm{n} = 5.55$ MHz/T the nuclear gyromagnetic ratio, $\hat{\mathcal{H}}_{Q^0}$ the nuclear quadrupole interaction in the neutral donor (in general different from that in the ionised case), resulting in a typical quadrupole splitting $f_{\rm q}^0\sim 10-100$~kHz \cite{fernandez2024navigating}, and $A$ is the Fermi contact hyperfine coupling, which in this specific donor takes the value $A \approx 98$~MHz. The hierarchy of energy scales $\gamma_{\rm e}B_0 \approx 39~\text{GHz} \gg A\approx 98~\text{MHz} \gg f_{\rm Q}^0 \approx 100 \text{kHz}$ ensures that the eigenstates of this Hamiltonian are to a good approximation the simple tensor products of the electron spin states $\{\ket{\downarrow},\ket{\uparrow}\}$ and the nuclear projections $\{\ket{m_I}\}$.
 
\begin{figure}[ht]
  \centering
  \includegraphics[width=14cm]{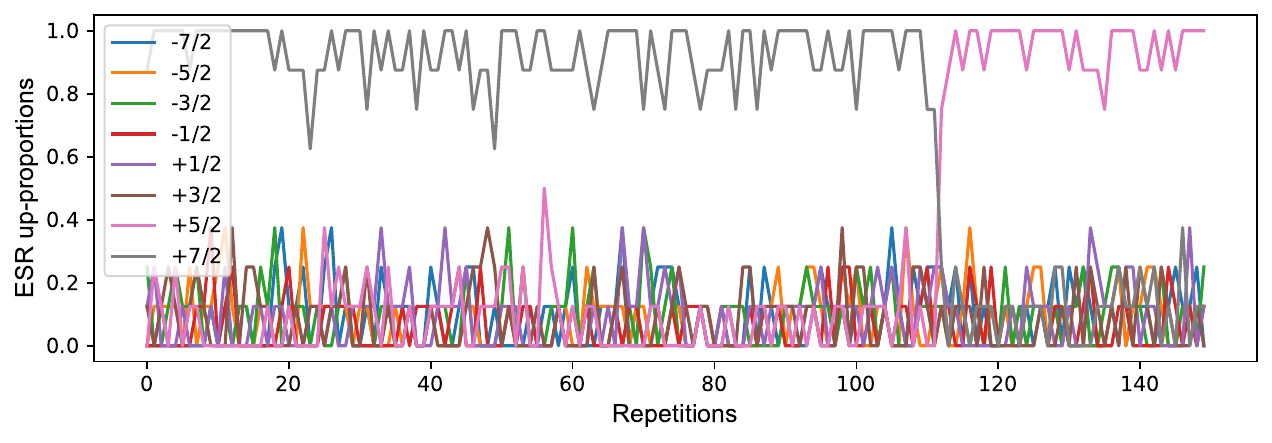}
  \caption{\textbf{Nuclear spin state readout.} The ESR frequency depends on the state of the nuclear spin, and hence the nuclear spin state can be determined by sequentially pulsing 8 ESR transitions. Each nuclear spin state is read by 10 single-shot adiabatic inversions and electron-spin readout measurements. At the optimal readout point, only one of the eight ESR resonances exhibits notably high electron spin-up probabilities. The nuclear spin state is then identified as the one that conditions the electron resonance with the highest up-proportion. }
  \label{fig:nuclear readout}
\end{figure}

Under these conditions, the electron exhibits 8 electron spin resonance (ESR) frequencies $f^{\rm ESR}_{m_I}$, separated by $\approx A$, dependent on the state of the nucleus. Starting from the electron $\ket{\downarrow}$, an adiabatic frequency sweep \cite{laucht2014high} around each of the $f^{\rm ESR}_{m_I}$ inverts the electron to the $\ket{\uparrow}$ state if the nucleus is in the state $\ket{m_I}$, i.e. performs a conditional quantum operation on the electron spin ancilla. The measurement is, to a good approximation, of quantum nondemolition (QND) nature \cite{pla2013high,joecker2024error} (see Section~\ref{sec:quantum_jumps} for deviations from QND), and can be repeated multiple times to increase the readout fidelity. Here we use 10 repetitions of the cycle [load $\ket{\downarrow}$ -- adiabatic ESR sweep around $f^{\rm ESR}_{m_I}$ -- measure electron state] for every nuclear orientation $m_I$. Because we are measuring a single nucleus, the electron spin responds at only 1 of the 8 possible resonances, as shown in Fig.~\ref{fig:nuclear readout}. Once the nuclear spin state $\ket{m_{I}}$ is determined, initialisation to other nuclear spin eigenstates can be achieved by using NMR $\pi$-pulses.
 
After preparing specific nuclear spin states, e.g. spin coherent state or cat states (Figs.~2, 3 and 4 of the main manuscript), the populations of each of the $\ket{m_I}$ states are extracted by repeating the preparation and nuclear readout cycles typically 50 times.
 
The electron spin readout and initialisation follows the standard method based on spin-dependent tunnelling \cite{Elzerman_2004,morello2010single}, whereby the electron can tunnel out of the donor if in the $\ket{\uparrow}$ state, whereas it remains bound to the donor if in the $\ket{\downarrow}$. The readout process automatically resets the electron spin in the $\ket{\downarrow}$ state. The method relies upon having a charge reservoir at very low temperature $T$, such that $\gamma_{\rm e}B_0 \gg k_{\rm B}T/h$, where $k_{\rm B}$ is the Boltzmann constant and $h$ is the Planck constant. The electron readout is, in fact, the only aspect of our experiment that demands operation at millikelvin temperatures. Here, the cold charge reservoir is embodied by the island of a single-electron transistor (SET), which also acts as the charge detector that signals the electron tunnelling event in real-time \cite{morello2010single}.
 
\subsection{Electron initialisation via a Maxwell's demon}
When using spin-dependent tunnelling for electron spin readout and initialisation \cite{Elzerman_2004,morello2010single}, the fidelity is inherently limited by the thermal broadening of the Fermi distribution in the charge reservoir \cite{Beenakker1991,Wiel2002}. 
The thermal limit to the fidelity of the $\ket{\downarrow}$ initialisation can be overcome by implementing a `Bayesian Maxwell's demon' \cite{Johnson2022}. This involves continuously monitoring the charge state of the donor and accepting it as $\ket{\downarrow}$ only after a stretch of time with no tunnelling events greatly exceeding the typical tunnel-out time of a $\ket{\uparrow}$ electron. If a tunneling event does occur, the timer restarts and the process loops until it succeeds, at which the $\ket{\downarrow}$ is captured by lowering its electrochemical potential. In this device, we accept a $\ket{\downarrow}$ electron when the timer is able to reach \SI{5}{\milli\second}.
 
\begin{figure}
    \centering
    \includegraphics[width=160mm]{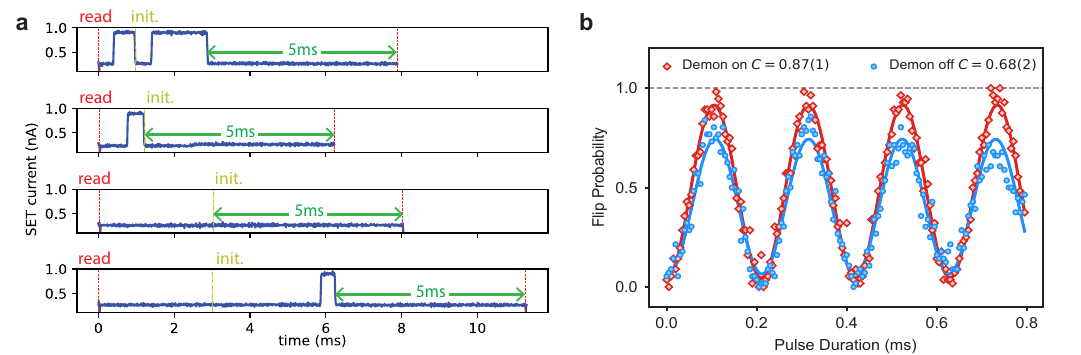}
    \caption{\textbf{Electron spin initialisation with Bayesian Maxwell’s demon}. 
    \textbf{a}, real-time traces of the SET current during the readout and initialisation stages. We allow a readout window of at most 3~ms to detect the tunnel-out event of an electron spin-up state $\ket{\uparrow}$, signalled by the current suddenly jumping to a high value. The initialisation stage starts immediately after another electron tunnels back to the donor (top two panels), switching the current back to a low value. If no tunneling event is registered within the 3~ms readout window (bottom two panels), the initialisation stage starts automatically. During the initialisation stage, an FPGA-based processor acts as the ``Demon", continuously monitoring tunnel-out events and cycling the process until no tunnel-out events are observed within a $t_{\text{init}} = 5$~ms window.
    \textbf{b}, Contrast of the neutral NMR Rabi oscillation when using (``Demon on") or not using (``Demon off") the Bayesian Maxwell’s demon to initialise the electron spin. 
    }
    \label{fig:mawell_demon}
\end{figure}

To assess the effectiveness of the Maxwell's demon initialisation we perform NMR Rabi oscillations while the donor is in its neutral state.
We drive the NMR pulse on-resonance with the transition between the states $\ket{-7/2, \downarrow}$ and $\ket{-5/2, \downarrow}$. 
Due to the hyperfine interaction, the resonance frequencies of the neutral NMR depend on the state of the electron \cite{pla2013high}. If a $\ket{\uparrow}$ electron is loaded onto the donor, an NMR pulse tuned to flip the nucleus when the electron is $\ket{\downarrow}$ will be completely off-resonance and leave the nucleus unaffected. Therefore, the contrast of the neutral-donor nuclear spin Rabi oscillation is a direct measure of the electron $\ket{\downarrow}$ initialisation fidelity. In Fig.~\ref{fig:mawell_demon} \textbf{b}, we demonstrate that the adoption of the Bayesian Mawell's demon increases the contrast from 0.68(2) to 0.87(1), which corresponds to reducing the $\ket{\downarrow}$ initialisation error by a factor $\approx 2.5$. 
 
\subsection{Nuclear quantum jumps - deviation from QND nuclear readout}
\label{sec:quantum_jumps}
 
\begin{figure}
  \centering
  \includegraphics[width=14cm]{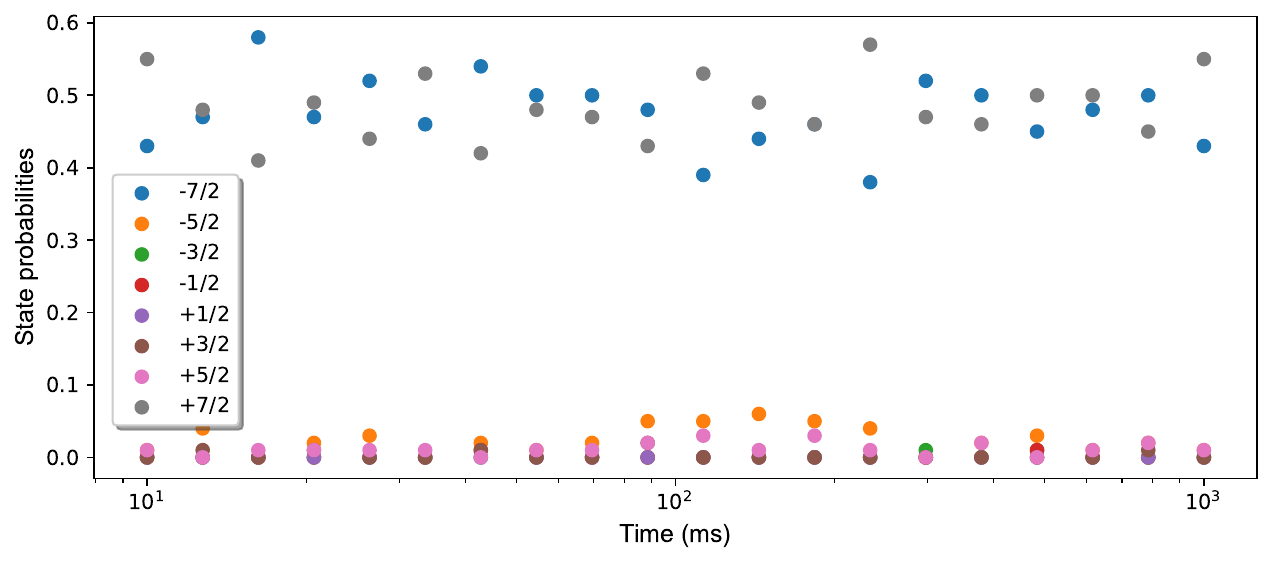}
    \caption{Nuclear spin populations for a $z$-oriented cat state, measured after introducing a variable wait time up to 1 second. The absence of e.g. a decay in the $\pm 7/2$ populations indicates that the intrinsic $T_1$ is much longer than the maximum wait time adopted here.}
    \label{fig:T1_cat}
\end{figure}

The energy relaxation time $T_1$ for nuclear spins in silicon is intrinsically very long, to be point of becoming immeasurable at cryogenic temperatures \cite{saeedi2013room,savytskyy2023electrically}, at least for the spin-1/2 $^{31}$P nucleus. As an initial confirmation that $T_1$ is very long also in $^{123}$Sb, we measured the state populations of a $\ket{\rm cat_{7/2}}_z$ state after a wait time of up to 1 second (Fig.~\ref{fig:T1_cat}). The data shows no clear trend of population change as a function of wait time, suggesting that $T_1$ is likely to be $\gg 10$~s, possibly orders of magnitude more.
 
Therefore, the only mechanism to affect the nuclear spin populations is the measurement, which requires the repetitive loading and unloading of the donor electron (Section~\ref{sec:GRvsPower}). The system Hamiltonian thus suddenly switches between $\mathcal{\hat{H}}_{D^+}$ (See Eq.~(1) in the main text) and $\mathcal{\hat{H}}_{D^0}$, where the latter includes the hyperfine interaction term $A\cdot \mathbf{\hat{S}}\cdot \mathbf{\hat{I}}$, and potentially a different quadrupole term $\mathcal{\hat{H}}_{Q^0}$. The slight change in the exact nuclear eigenstates between the two Hamiltonians introduces the possibility of `quantum jumps' \cite{pla2013high}, which can be viewed as the consequence of the measurement not being exactly QND \cite{joecker2024error}. The process whereby electron tunnelling causes nuclear spin flips is sometimes called `ionisation shock'~ \cite{hile2018addressable}. In our system, this process constitutes effectively the main source of state preparation and measurement (SPAM) error.
 
\begin{figure}
  \centering
  \includegraphics[width=12cm]{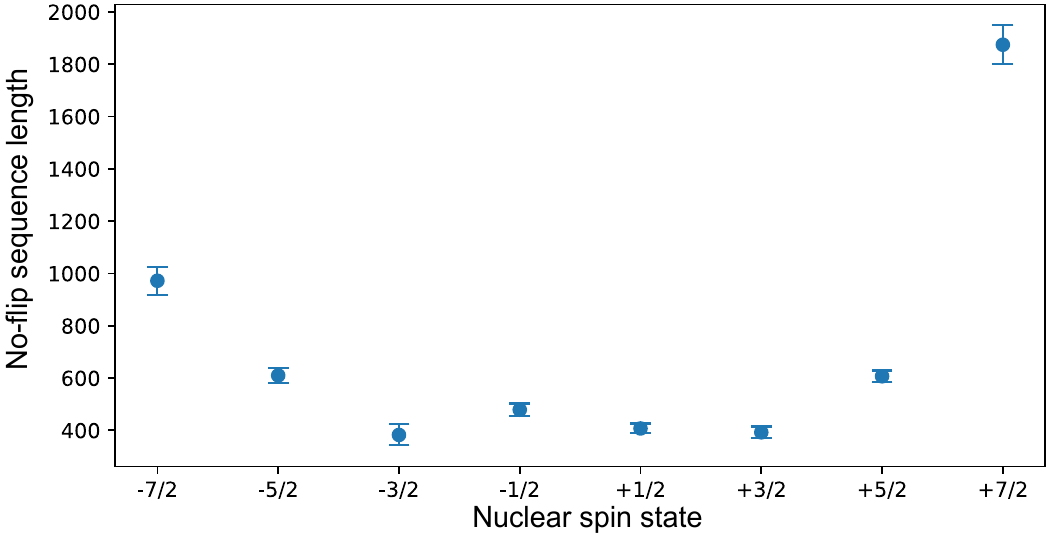}
  \caption{
  \textbf{Nuclear quantum jumps caused by ionization shock}.
  For each nuclear spin state $\ket{7/2,m_{I}}$, we repeat single-shot readout of the electron with an ESR pulse at the frequency of $f^{\rm ESR}_{m_I}$ until 10 consecutive failure detections of electron spin-up $\ket{\uparrow}$ occur. Upon such an event, the nuclear jump event is registered, following which we initialise the nuclear state back to $\ket{7/2,m{I}}$ by NMR pulse and repeat the process. Each "no-flip sequence length" value is obtained from fitting an exponential decay to a histogram of the occurrence of sequences without quantum jumps against the sequence length.
  }
  \label{fig:nuclear_jumps}
\end{figure}

We measure the susceptibility of each nuclear eigenstate $\ket{m_I}$ to ionization shock by performing repeated nuclear readouts on the state, and plotting the typical length of sequences where no spin flips occur (Fig.~\ref{fig:nuclear_jumps}). We observe a clear trend where states with lower $|m_I|$, i.e. close to the equator of the spin-7/2 Bloch sphere, display a higher probability of quantum jumps. A detailed model of this trend is left to future work. 
 
\section{Other devices}
\label{sec:other_devices}
 
\begin{figure}[ht!]
    \centering
    \includegraphics{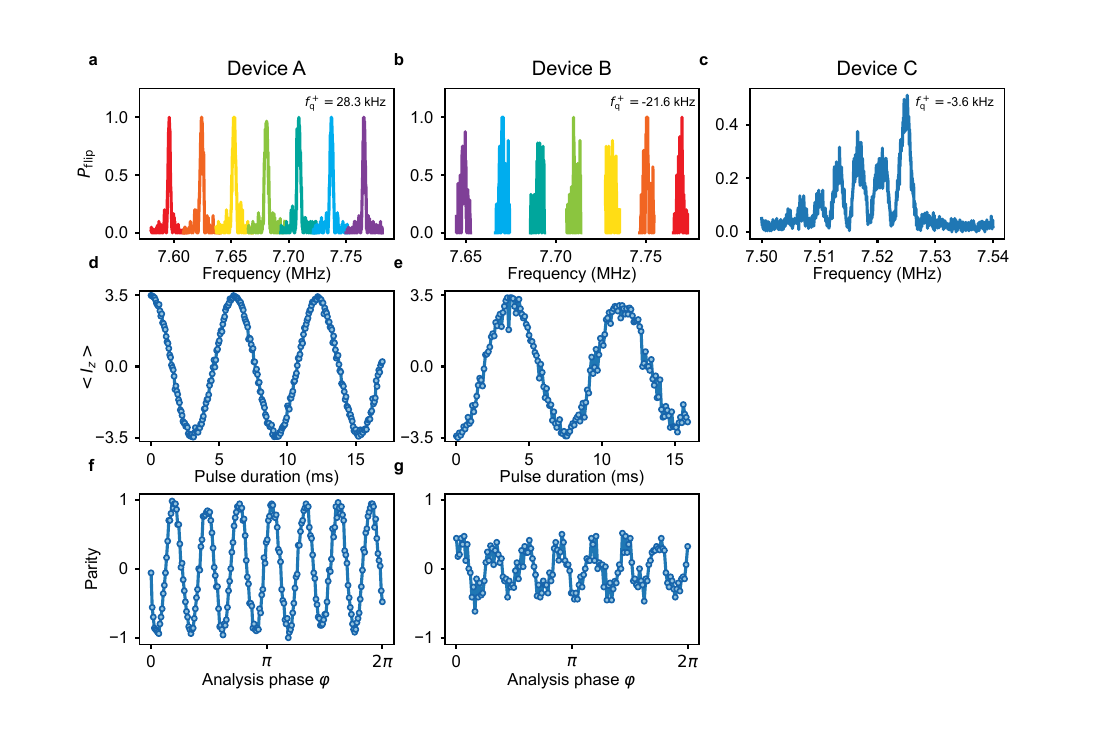}
    \caption{\textbf{NMR spectra, covariant SU(2) rotations and parity oscillations in other devices.} \textbf{a-c}, NMR spectra displaying the 7 NMR transitions of the $^{123}\textrm{Sb}$ nucleus. The spacing between the peaks indicates the quadrupole splitting $f_q^+$, as denoted in the plot. Different colors represent different datasets, where the color labelling is consistent with Fig.~1 of the main text (i.e. red represents $f_1$, purple $f_7$). In \textbf{a-b}, measurements are performed after initialisation in one of the nuclear spin states involved in the respective NMR transition. In \textbf{c}, the NMR spectrum is a single dataset without nuclear state initialisation, hence the irregular peak heights.
    \textbf{d-e}, Covariant SU(2) rotations on devices A and B. Device B shows a significant decay in the covariant SU(2) rotation shown in panel \textbf{e}.
    \textbf{f-g}, Parity oscillations of a $\ket{\textrm{cat}_{7/2}}_z$ state for devices A and B. Due to operational limitations caused by the small quadrupole splitting in device C, no covariant SU(2) rotations and parity oscillations were performed for that device.}
    \label{fig:other devices}
\end{figure}

For this project, we fabricated a batch of $^{123}\textrm{Sb}$-implanted devices devices, three of which were measured in a dilution refrigerator. 
We label the devices `A' (the one used for all data shown in this paper), `B' and `C'. Devices A and C were produced with the same $^{123}\textrm{Sb}$ implantation parameters (18~keV, $5\times10^{11}$~cm$^{-2}$), while device B was produced with different $^{123}\textrm{Sb}$ implantation parameters (10~keV, $4\times10^{11}$~cm$^{-2}$). All other aspects of devices A-C are nominally identical. Fig.~\ref{fig:other devices} shows the NMR spectra, covariant SU(2) rotations and parity oscillations performed  on these devices. The NMR spectra of the three different devices, shown in Fig.~\ref{fig:other devices}a-c, show a wide range of values for the quadrupole splitting $f_{\rm q}^+$, where even the sign can vary from device to device. The sign and magnitude of $f_{\rm q}^+$ depends on the orientation of the applied magnetic field $B_0$ with respect to the principal axes of the electric field gradient at the nucleus.
\cite{asaad2020coherent, FrankePRL2015}. In Device C, this conjures a value of $f_{\rm q}^+ =-3.6$~kHz (Fig.~\ref{fig:other devices}c, device C), which complicates the execution of even simple Givens rotations or covariant SU(2) rotations due to the low power necessary to resolve individual transitions.
 
Device A showed notably superior performance to the other two devices. In device B, we observed a beating pattern in Ramsey experiments between all transitions, presumably due to coupled $^{29}\textrm{Si}$ nuclei in the vicinity of the donor. This lead to imperfect SU(2) rotations and, consequently, low-contrast parity oscillations. Using silicon materials with 730~ppm residual $^{29}$Si isotopes, we typically expect of order 2 or 3 significantly hyperfine-coupled $^{29}$Si nuclei. Future experiments will seek to incorporate $^{28}$Si materials with more extreme enrichment, down to $\sim 2$~ppm residual $^{29}$Si \cite{acharya2023highly}, where the occurrence of a donor with no coupled $^{29}$Si nuclei will become commonplace.
 
\newpage
\noindent
 
\newpage
% \bibliographystyle{naturemag}
% \bibliography{bibliography.bib}% common bib file
\providecommand{\noopsort}[1]{}\providecommand{\singleletter}[1]{#1}%

\end{document}